# **Title:** Improving outdoor navigation for people with blindness using an AI-driven smartphone application and personalized audio guidance

**Authors:** Raymond Liu[1] and Patrick Slade[1,2]*

**Affiliations:**
[1]John A. Paulson School of Engineering and Applied Sciences, Harvard University; Boston, MA, USA.
[2]Kempner Institute for the Study of Artificial and Natural Intelligence, Harvard University; Boston, MA, USA.
*Corresponding author. Email: slade@seas.harvard.edu

**Summary:** Globally, 340 million people have blindness or moderate to severe visual impairment (BVI)[1] which limits independent outdoor navigation[2] and negatively affects their health and quality of life[3,4]. We surveyed 112 people with BVI and found that an ideal outdoor navigation aid must be able to perform turn-by-turn directions, path guidance, and obstacle detection and avoidance. Existing navigation tools such as white canes, guide dogs, and electronic travel aids often lack one or more of these criteria and may be expensive or inaccessible[5,6]. Here we introduce Mobilio, a smartphone application that incorporates machine learning, sensor fusion algorithms, and personalized audio feedback to meet all of the outdoor navigation criteria. The reliability of the smartphone sensors and models used for navigation were assessed with engineering tests in representative navigation scenarios. We performed a series of experiments where Mobilio personalized audio feedback for participants with BVI ($n$ = 14), guided them along an outdoor community path, and helped them navigate an obstacle course. Participants walking with Mobilio and a white cane reduced time to navigate a community path by 13 ± 3% and environmental contacts by 41 ± 5% compared to using Google Maps and a white cane. Mobilio achieved similar outdoor navigation reliability as a human guide. Participant surveys reported that Mobilio was easy to use, had a low perceived workload, and provided intuitive audio feedback. This work provides an accessible and personalized tool that may be an effective outdoor navigation aid to increase independence for people with BVI.

**Main Text:** People with BVI face many navigation challenges that limit independence and negatively affect health and quality of life. Mobility is a crucial need for daily living but is severely limited by BVI, which is associated with a decreased ability to independently navigate, decline in movement and physical health, and greater risk of collisions or falls[2,4]. BVI is also associated with an increased risk of depression, a reduced emotional well-being, and a lower quality of life[3,4]. People with BVI often require assistance with performing routine activities requiring navigation such as grocery shopping, attending school, or working, limiting independence and contributing to a lower educational status[7] and reduced income or unemployment[8]. Accessible assistive technology that helps people with BVI navigate independently could improve physical and mental health, socioeconomic status, and quality of life.

Surveys from 112 people with BVI indicated that their ideal navigation aid must meet a critical set of navigation criteria: outdoor turn-by-turn directions, path guidance, and obstacle avoidance (Extended Data Table 1). Outdoor turn-by-turn directions provide global pathing to enable users to travel along a route to a specific destination, particularly routes that the user has not experienced before[9]. Path guidance provides local pathing to help users stay oriented while navigating along a desired route and can be done either with continuous guidance or by periodically identifying nearby landmarks[10]. Obstacle avoidance helps users avoid collisions with objects in their path which can cause injuries that often result in medical consequences[11]. Surveyed users also indicated a preference for navigation aids to be small and lightweight (Extended Data Table 1), as well as low cost (Extended Data Table 2). Past studies suggest similar navigation and design criteria are important for navigation aids[5,9,10].

White canes and guide dogs are conventional solutions that are useful for obstacle detection, but neither can navigate along unfamiliar routes. White canes are a simple and low-cost method for obtaining haptic information about the ground surface and nearby obstacles[12,13] and act as an indicator that the user has BVI[14], but cannot detect obstacles beyond the length of the cane. Guide dogs can detect obstacles and provide path guidance but cannot provide turn-by-turn directions to navigate to a new destination[5]. Guide dogs are only used by approximately 2% of people with BVI[6] because they are expensive and time-consuming to train and care for[15].

Electronic travel aids (ETAs) and electronic orientation aids (EOAs) use technology to improve navigation for people with BVI, but they are typically designed to address a subset of the navigation criteria and few are readily accessible. ETAs help users avoid obstacles while EOAs guide users to desired destinations with turn-by-turn directions or by providing information on nearby landmarks and points of interest[10]; here we refer to all electronic aids as ETAs. Outdoor turn-by-turn directions are possible by using a Global Positioning System (GPS)[16] receiver to track the user's position. Computer vision enables some ETAs to provide orientation information by identifying key objects and landmarks in the environment[17]. For obstacle detection, some ETAs measure distance using ultrasonic, infrared, or light detection and ranging (LIDAR) sensors[18] that are either handheld[19], mounted to a white cane[20,21], or worn on the user's body[22]. A few ETAs fuse multiple sensing modalities to meet more than one navigation criteria but they cost hundreds to thousands of dollars[20,23,24], do not meet user-specified design criteria such as size and weight[5,20,25] (Extended Data Table 1), and are unlikely to be widely distributed or replicated[5,26–28]. ETAs are rarely experimentally evaluated to determine whether they improve mobility metrics such as walking time, safety, or user confidence[5,27]. Some evaluations are limited by having few participants, only involving blindfolded people without BVI, or not comparing performance with a white cane[5,27].

Smartphones are a promising platform for a navigation aid, but existing smartphone applications do not meet all of the navigation criteria. Smartphones are small, lightweight, widely used[29], contain a variety of sensors, and support easy deployment of software. Navigation applications like Google Maps[30] and Microsoft Soundscape[31] provide GPS-based turn-by-turn directions or information on nearby landmarks and are commonly used by people with BVI[32] (Extended Data Table 3). However, smartphone GPS measurements have reported average errors of 5 meters under open sky[33] and compass measurements can be inaccurate due to magnetic interference, which may limit the precision of the directions and reduces usability for people with BVI. Some systems use pre-placed Bluetooth beacons and floor plans[34–36], floor markers[37], or digital mappings[38–40] for navigating known areas with higher precision than GPS, but these approaches are not easily generalizable for widespread use. Visual interpreter tools[41–43] provide scene descriptions from a sighted human or visual language model which can provide path guidance, but require human assistants and constant internet access. Smartphone-based approaches to obstacle detection use computer vision methods but often rely on external sensing or processing capabilities[44]. Computer vision methods on smartphones are limited by a smartphone's processing power, field-of-view, and potential motion artifacts from the phone moving during walking. Few smartphone-based approaches are experimentally evaluated to assess navigation benefits for people with BVI[17,44].

Personalizing guidance from an ETA may enhance navigation performance. Systematic personalization of assistance has improved device benefits across many different assistive devices and patient populations[45,46] and may improve mobility metrics in navigation aids[20,35,36]. People with BVI are heterogeneous in types of impairment and in acquisition of spatial knowledge[47], indicating that optimal feedback could vary widely for each person[45,46].

In this work we introduce Mobilio, a smartphone application that provides spatial audio feedback while performing turn-by-turn directions, path guidance, and obstacle avoidance to improve outdoor navigation for people with BVI. The smartphone is strapped to the user's chest (Fig. 1a-b) and is meant to complement conventional tools such as a white cane that provide environmental safety and physical feedback. The user's movement is estimated by performing visual-inertial odometry (VIO) using data from the camera and inertial measurement unit. Mobilio provides turn-by-turn directions between GPS waypoints (Fig. 1c) and corrects errors in GPS and compass readings by incorporating VIO. Mobilio provides path guidance along walkable areas by fusing VIO with outputs from a machine learning model[48] that we trained to segment key elements of street environments[49] for mapping nearby surfaces (Fig. 1d and Extended Data Fig. 1a). Obstacles are detected by fusing VIO with depth images to create a 3D occupancy grid map[50], which is used to guide the user to avoid collisions (Fig. 1e and Extended Data Fig. 1b). All algorithms and models run on the smartphone in real time. Additional details are provided in the methods section titled "Navigation algorithms". Mobilio is a uniquely capable navigation assistant that addresses all major navigation and design criteria, providing a scientific tool to investigate how to best assist users with BVI. We evaluated the navigation capabilities of Mobilio through a series of experiments where participants with BVI had audio feedback personalized to them during outdoor walking, navigated an outdoor community path, and navigated an obstacle course (Supplementary Video 1 and 2).

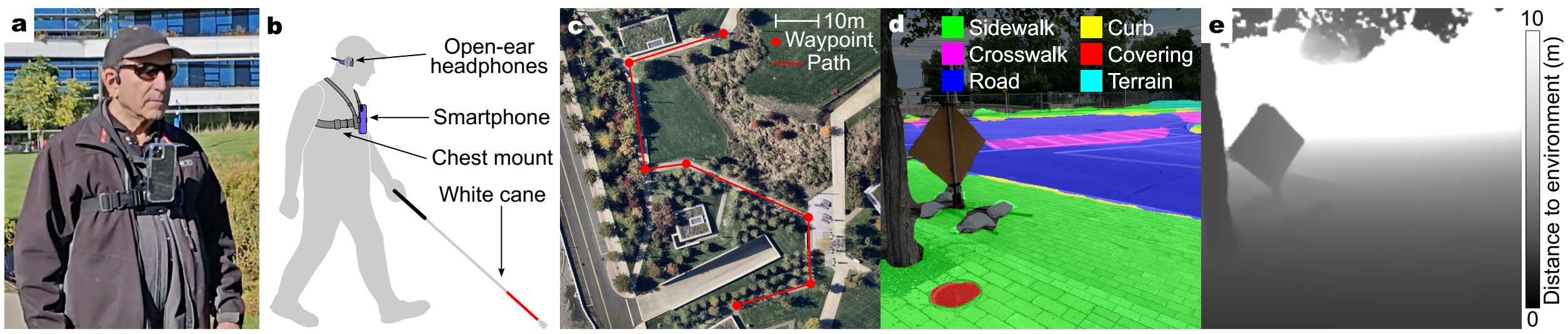

**Fig. 1. Overview of Mobilio. a,** A participant walking in a community setting while wearing a smartphone with the Mobilio application. **b,** During navigation, participants use an iPhone 12 Pro smartphone strapped to their chest, open-ear headphones to provide spatial audio feedback, and a standard white cane. The iPhone 12 Pro contains a variety of sensors including a camera, an inertial measurement unit, GPS receiver, and 3D LIDAR sensor. **c,** The spatial audio feedback provides turn-by-turn directions to guide users along GPS waypoints towards an outdoor destination. **d,** Path guidance along walkable paths is provided by creating a 2D grid map by fusing visual-inertial odometry (VIO) with key elements of street environments labeled by a semantic segmentation model. **e,** Obstacles are detected by fusing VIO with depth images from the camera and LIDAR to create a 3D occupancy grid map of the user's surroundings. All models and computation are performed on the smartphone in real time.

## Results and Discussion

Sensor and algorithm performance

We evaluated Mobilio's sensing and algorithmic performance during scene segmentation, obstacle distance estimation, and visual-inertial odometry with separate experiments. Mobilio's segmentation model had similar validation performance on key street scene surfaces compared to a large state-of-the-art pre-trained model requiring 250 times the compute, providing high performance segmentation while accommodating smartphone computation limits (Fig. 2a). The segmentation outputs for several representative images indicate the model can account for some weather, paint degradation, and low light conditions (Fig. 2b and Extended Data Fig. 2a), but extreme versions of these conditions result in poor performance (Fig. 2c and Extended Data Fig. 2b). Motion artifacts from camera blur during walking (Supplementary Video 2) were visually much less than the threshold of image blur necessary to degrade segmentation model performance (Extended Data Fig. 3). A participant walked towards common navigation obstacles in a motion capture laboratory to evaluate obstacle distance estimation performance of the smartphone LIDAR or the camera using a monocular depth estimation model, Depth Anything V2[51] (DAV2) (Fig. 2d). LIDAR and DAV2 provided consistent estimates of the distance to the obstacles compared to the ground-truth motion capture distance with percentage errors[52] of 5.7% and 14.6%, respectively (Fig. 2e). LIDAR and DAV2 had similar percentage errors across all of the representative obstacles, indicating that both methods could provide sufficient depth estimation for obstacle detection during navigation (Fig. 2f). A participant walked 300 meters back and forth along a community outdoor path (Fig. 2g) to determine the VIO position estimate had a mean error of 1.2 meters, or 0.4% error (Fig. 2h). This supports prior studies that indicate smartphone VIO can reliably estimate the user's position during hundreds of meters of movement[53].

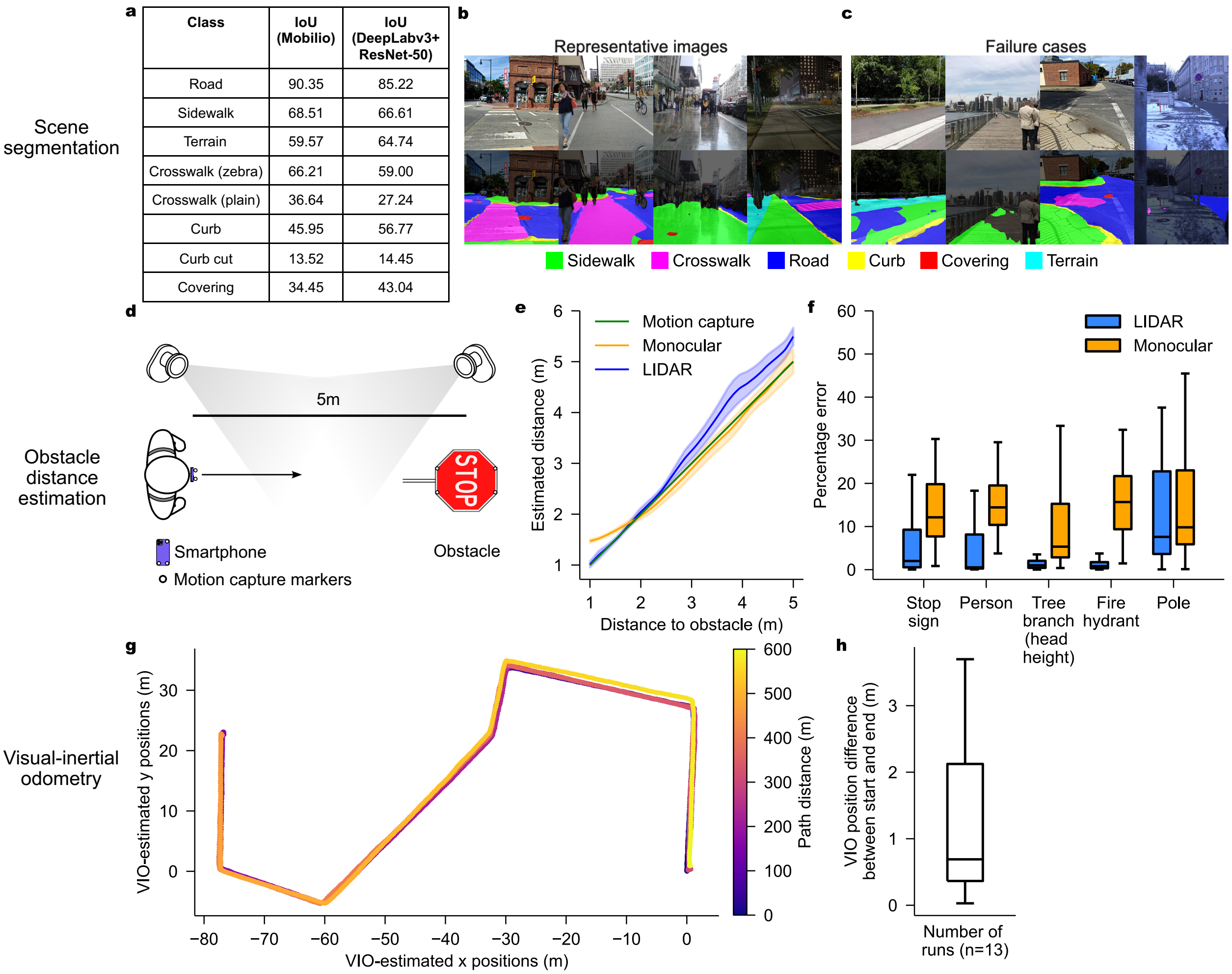

**Fig. 2. Technical validation of Mobilio. a,** Mobilio's lightweight scene segmentation model achieves similar intersection-over-union performance, representing the portion of the image class that was segmented correctly, on key street environment surfaces in comparison with a large state-of-the-art model pre-trained on the same dataset. **b,** Representative images from the segmentation model indicate the model can distinguish the environment even in conditions with degraded crosswalk paint, rain, and low light. **c,** The segmentation model may perform poorly on sidewalks that appear perceptually similar to roads, on walkable paths like a pier that are not a sidewalk, on crosswalks where the painted lines are faded, or during weather like snow that visually occludes the terrain. **d,** Smartphone depth estimation was evaluated in an indoor motion capture space. Motion capture tracked the position of the smartphone as the person approached five common obstacles, such as a stop sign. Each obstacle was approached five times from 5 meters away. **e,** Distances to objects were measured in real time on the smartphone using LIDAR and Depth Anything V2[51] (DAV2), a monocular depth estimation model. LIDAR and DAV2 consistently estimated the distance to the objects compared to ground-truth measures from motion capture. **f,** LIDAR and DAV2 had similar errors for each object, with an average error across objects of 5.7% and 14.6%, respectively. **g,** The reliability of VIO position estimates were evaluated as a participant walked back and forth 300-meters along a community path. **h,** The average error between the start and end points of the path was 1.2 meters or 0.4% error for 13 completions of the path.

Audio feedback parameters and personalization

Mobilio provides spatial audio feedback through open-ear headphones to guide users to turn towards a desired heading by playing a beeping sound at a fixed rate. Open-ear headphones are suitable for providing spatial audio cues while allowing users to hear other sounds in the environment. Audio feedback is provided spatially such that the beeps appear to be coming from the direction of the desired heading. For example, beeps are played more prominently in the user's left ear when the user should turn left. Three parameters define the audio feedback: audio rate, pitch range, and scaling factor (Fig. 3). The audio rate determines how fast the beeping sounds are played. The pitch range determines the lowest possible pitch that plays when the user has the greatest absolute heading error, which is defined as the error between the current and desired headings. For example, the audio pitch of each beep increases as the user turns towards the desired heading and reaches its maximum value when the user is directly facing the desired heading. The scaling factor is multiplied with the user's heading error to over- or under-exaggerate the spatial audio direction and pitch.

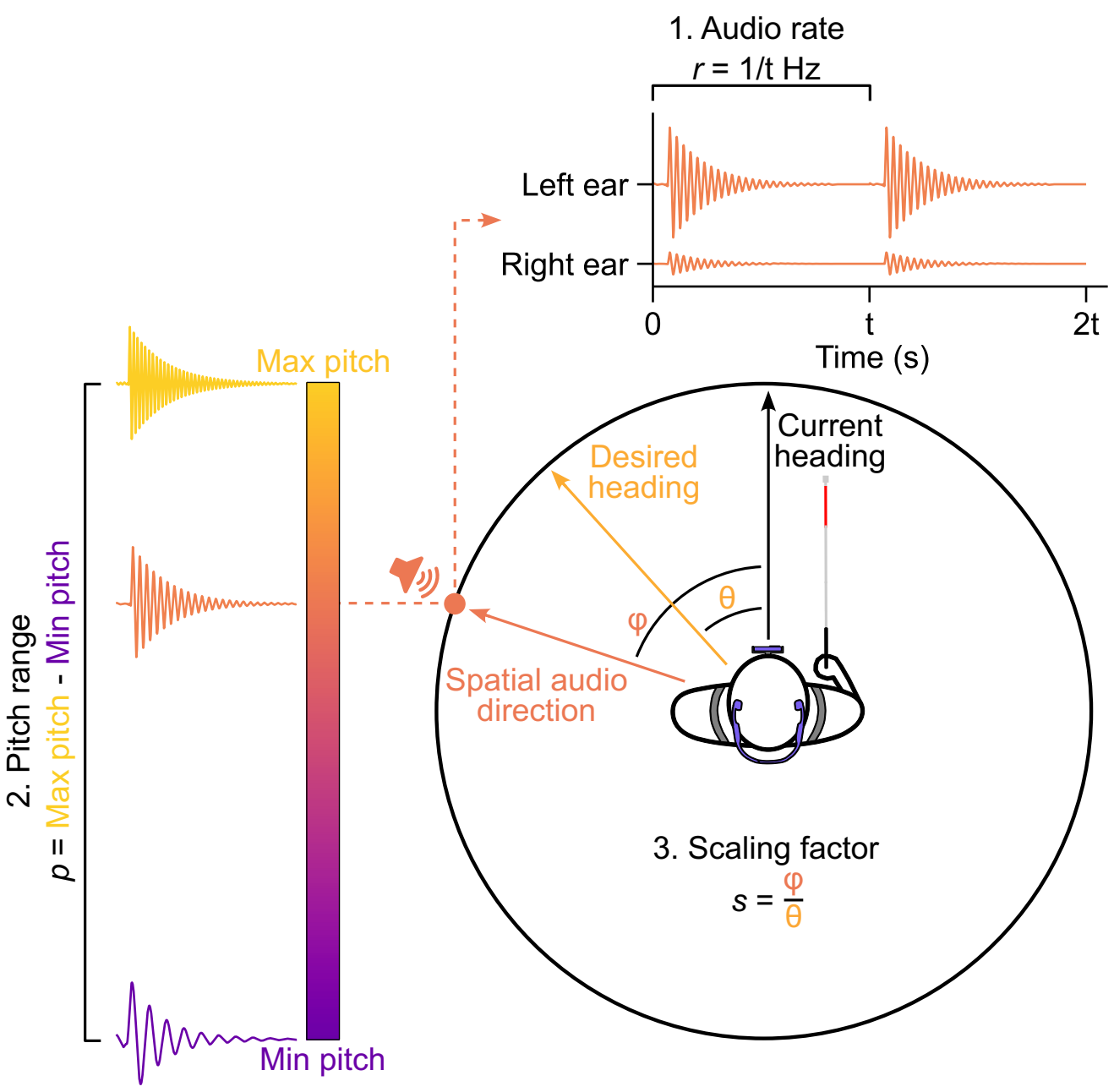

**Fig. 3. Spatial audio feedback system.** The spatial audio feedback that guides users during navigation is defined by three parameters. The audio rate ($r$) defines a constant rate at which the beeping sounds are played. The pitch range ($p$) defines how the heading error is mapped between a minimum and maximum audio pitch. The scaling factor ($s$) multiplies the user's heading error to over- or under-exaggerate the spatial audio direction.

Audio feedback was personalized for each user using an approach called human-in-the-loop optimization to reduce their heading error during outdoor walking. Mobilio optimizes audio feedback parameters during walking using covariance matrix adaptation evolution strategy (CMA-ES)[54] by systematically tuning the three audio parameters to reduce heading error when turning towards a desired heading (Fig. 4a). During optimization, the user walks and receives audio feedback from one set of parameters during each turn towards a desired heading. Each parameter set is scored with the normalized average heading error during the first five seconds of a turn (Extended Data Fig. 4). Each generation of optimization consists of seven parameter sets that are ranked by error and used to update the estimate of the optimal parameters. The updated optimal parameter estimate is sampled, selecting a new generation of seven parameter sets to evaluate. The generic parameters used to initialize optimization were selected through hand-tuning by a pilot participant. Participants ($n$ = 14; 7 men and 7 women; age, 45.4 ± 13.2 yr; body mass, 82.7 ± 24.2 kg; height, 1.69 ± 0.11 m; 9 blind and 5 visually impaired) performed 196 turns, or 28 generations of optimization, to reach randomly sampled target locations in an unobstructed 10-meter radius area (Extended Data Fig. 5). A new desired heading between 60 and 120 degrees to the left or the right was provided every 5 to 8 seconds to evaluate many sets of audio parameters during 22 minutes of walking. The optimizer's convergence parameter decreased throughout the optimization, indicating decreasing uncertainty in the estimate of the optimal parameter values (Fig. 4b). User heading error decreased by 19 ± 3% (ANOVA, $n$ = 14, $P$ = 6.5e-05) during validation tests where participants completed 32 turns with both the generic starting parameters and the optimized parameters (Fig. 4c). Optimized parameters were different across participants, supporting the idea that personalization may be increasingly important for populations like people with BVI due to the unique needs of each participant (Fig. 4d). Continuous audio feedback from Mobilio was similarly effective in preventing over-rotating when performing turns compared to previous methods, which is detailed in the methods section titled "Spatial audio feedback".

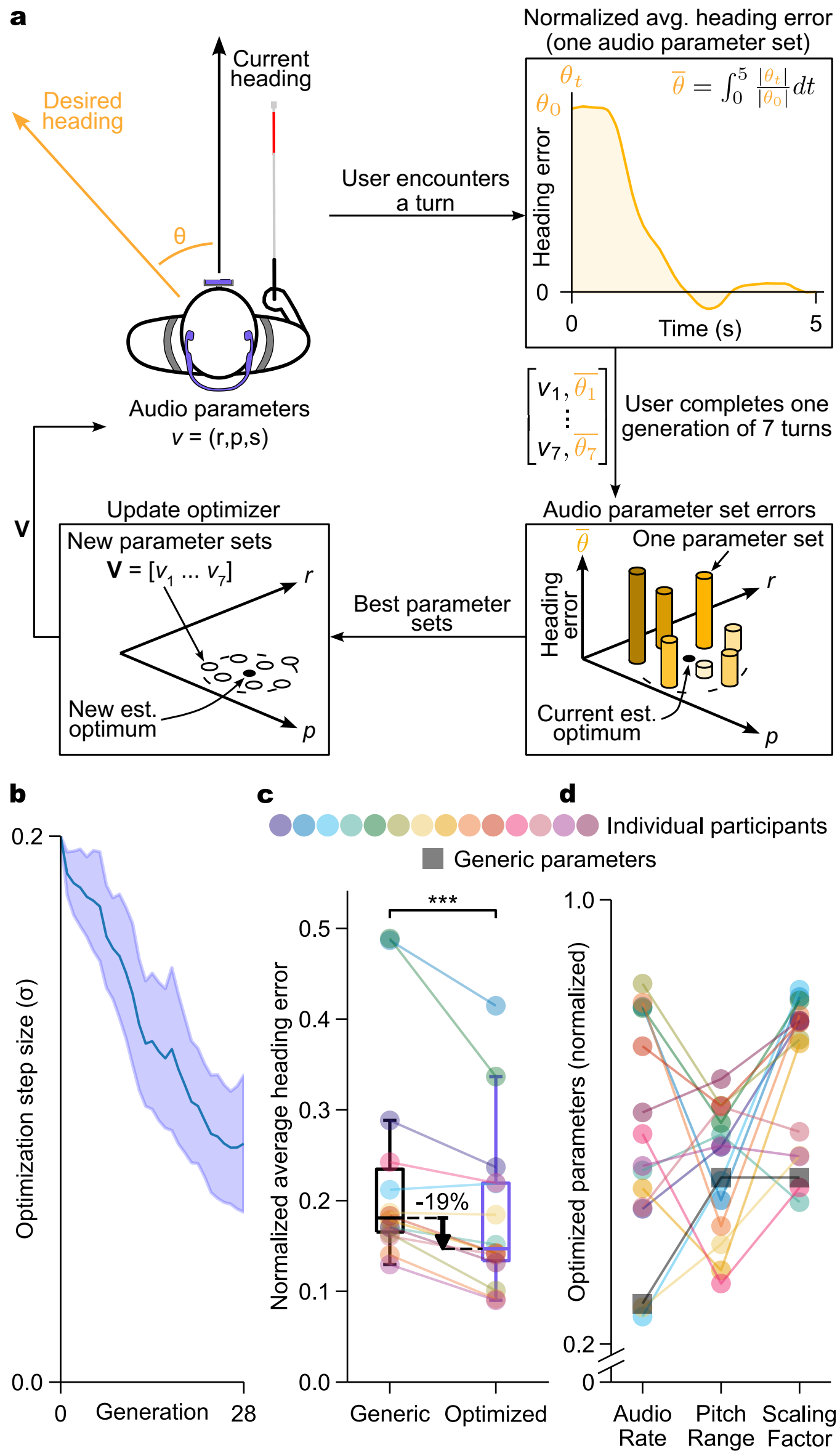

**Fig. 4. Personalizing audio feedback to improve navigation. a,** The goal of this human-in-the-loop optimization is to identify audio feedback parameters that minimize the user's heading error as they perform turns during walking. During optimization, the participant walks with a white cane and Mobilio while receiving spatial audio feedback to complete turns towards a series of desired headings. The audio feedback parameters are unique for each turn, with 7 audio parameter sets and their corresponding turns comprising one generation of optimization. The normalized average heading error over the first five seconds of each turn provides the score for the current parameter set. The optimizer ranks each parameter set in the generation based on this normalized average heading error. The heading error across two parameters ($p$, $r$) is illustrated for ease of visualization, but all three parameters are optimized. The CMA-ES optimizer is updated with the best performing parameter sets and a new generation of parameter sets is chosen to evaluate. **b,** The optimizer convergence parameter (σ) continually improved during 22 minutes of optimization ($n$ = 14). The error band represents the 95% confidence interval of the mean. **c,** During 7-minute validation tests, optimized parameters reduced heading error compared with the generic starting parameters. Boxes extend from the lower to upper quartile values of the data with a line at the median. Whiskers extend between the minimum and maximum of the data values within 1.5 times the interquartile range. **d,** Optimized parameters were diverse across different participants.

Navigation experiments for participants with BVI

Mobilio improved outdoor navigation in a community setting by reducing the number of times participants needed human intervention and decreasing travel time compared to Google Maps. We selected a path in a busy community outdoor setting which included other pedestrians, cars driving by, construction on neighboring buildings, and maintenance crews performing noisy tasks like mowing the grass (Supplementary Video 1 and 3). Participants walked with a white cane (Fig. 5a) along a 150-meter community path (Fig. 5b) in three conditions: a human guide providing verbal turn instructions representing ground truth guidance, Mobilio, and Google Maps. The conditions were randomized and presented in a double-reversal order, as ABCCBA, for a total of 6 runs with the starting and ending points alternating between runs to make it more difficult for the participant to remember the series of turns (Extended Data Table 4). Participants were corrected with verbal instructions when they asked for assistance, stopped for 10 seconds, or walked the wrong way for 10 seconds. Minimizing the number of corrections with verbal instructions is important because pedestrians may not be nearby to help a person with BVI who is navigating independently. Participants needed 72 ± 14% fewer corrections per run with Mobilio compared to Google maps (repeated measures analysis of variance (ANOVA), $n = 14$, $P = 1.8e\text{-}3$) and needed 82 ± 16% fewer corrections with a human guide compared to Google Maps (ANOVA, $n = 14$, $P = 1.8e\text{-}4$) (Fig. 5c and Extended Data Fig. 6). Participants took an average of 13 ± 3% less time to traverse the path when using Mobilio compared to Google Maps (ANOVA, $n = 14$, $P = 2.5e\text{-}3$) and took 13 ± 2% less time with a human guide compared to Mobilio (ANOVA, $n = 14$, $P = 5.1e\text{-}4$) (Fig. 5d). The ordering of the assistance methods in experiments did not have a significant effect on the average number of corrections (ANOVA, $n = 14$, $P = 0.58$) or on path completion time (ANOVA, $n = 14$, $P = 0.94$).

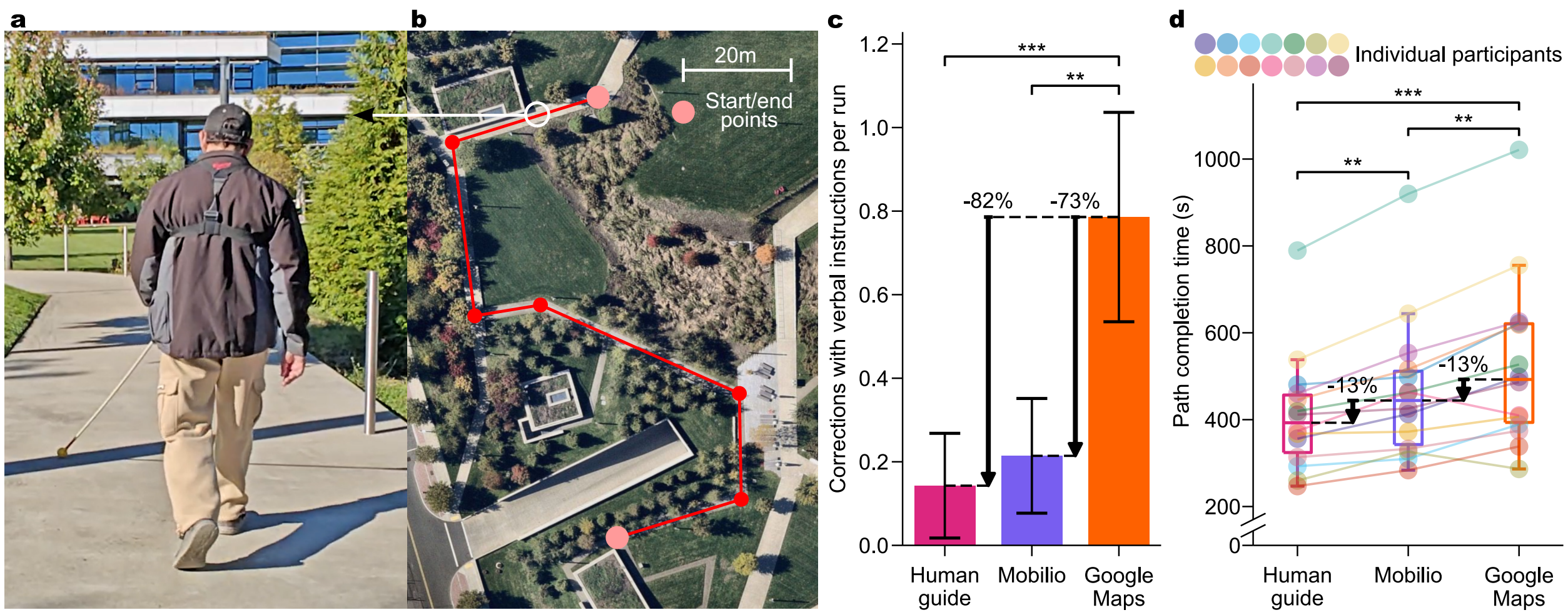

**Fig. 5. Outdoor navigation experiment. a,** A participant walking along the community path during the experiment. **b,** Map of the 150m community path with red markers indicating turns. The start and end points alternated each time the participant completed the path. **c,** Participants received a similar number of corrections with verbal instructions during the human guide and Mobilio conditions (ANOVA, $n = 14$, $P = 1.0$) and significantly more corrections during the Google Maps condition. Error bars represent one standard error of the mean. **d,** Participants walked the path while using a white cane during each of three navigation assisted conditions in a randomized, double-reversal order: verbal turn instructions from a human guide, the Mobilio smartphone application, or Google Maps. Participants had a lower path completion time when using Mobilio compared to Google Maps (ANOVA, $n = 14$, $P = 2.5e\text{-}3$). Boxes extend from the lower to upper quartile values of the data with a line at the median. Whiskers extend between the minimum and maximum of the data values within 1.5 times the interquartile range.

Mobilio provided continuous audio feedback from sensor fusion algorithms to achieve a similar turn-by-turn reliability to a human guide during this real-world navigation experiment. Achieving human level guidance supports the importance of using multi-modal sensing to provide robust navigation assistance. Google Maps required significantly more verbal corrections than the other methods, which the research team observed may be due to early and poorly timed turning directions, as well as a lack of continuous feedback. Additional details are provided in the methods section titled "Community path navigation". Navigation time was reduced by 13% with Mobilio and 27% with a human guide compared to Google Maps which are relatively minor improvements, but the evaluated path was relatively simple with few obstacles directly on the path and with distinct edges for easy shorelining. During longer and more complex navigation scenarios, Mobilio may provide a greater reduction in navigation time by reducing the number of times participants get confused or lost on the route. Mobilio's reduced navigation time and human-level turn-by-turn reliability support that it may be an effective tool to help people BVI navigate outdoors.

During obstacle course navigation, Mobilio reduced the number of times the cane or person's body contacted the environment, but increased travel time compared to using only a white cane. Participants navigated through a hallway that was 1.8m wide with two 0.85m wide obstacles placed randomly at one of the predetermined placement intervals (Extended Data Fig. 7a). Participants navigated the hallway in two conditions, using either a white cane or Mobilio and a white cane. The conditions were randomized with a double reversal ordering, as ABBA (Extended Data Table 5). In each condition, participants navigated the hallway 10 times with different obstacle locations for each run. Participants using Mobilio reduced environmental contacts by 41 ± 5% (Extended Data Fig. 7b) and increased hallway completion time by 17 ± 5% (Extended Data Fig. 7c) compared to no assistance. The reduction in environmental collisions during navigation suggests that Mobilio may increase safety for users with BVI when navigating around obstacles. The added cognitive load of interpreting the audio feedback from Mobilio likely contributes to the increased completion time. Mobilio may help people with BVI avoid injuries due to collisions or falls by navigating around obstacles.

Participant surveys and feedback

Post-experiment participant surveys reported that Mobilio provided intuitive audio feedback and had a high usability and low workload compared to a white cane or Google Maps. Participants evaluated the ease of use and effectiveness of Mobilio with an average score of 79 ± 7 on the System Usability Scale survey[55] (Extended Data Table 6). Mobilio's usability score is in the 84th percentile among hundreds of aggregated usability studies[56] and is substantially higher than scores of 57 (23rd percentile) for an augmented white cane and 65 (41st percentile) for a white cane in a previous study of participants completing similar tasks[20]. Participants completed the NASA Task Load Index survey[57] to evaluate the perceived workload of Mobilio, a standard white cane, and Google Maps with average scores of 26 ± 12, 27 ± 14, and 45 ± 14 (Extended Data Table 7), corresponding to the 10th, 12th, and 40th percentiles in a distribution of over one thousand workload scores, respectively[58]. A lower score and percentile correspond to a lower workload. Mobilio's perceived workload is similarly low compared to the perceived workload of a white cane. The decrease in completion time during outdoor navigation and the increase in completion time during hallway obstacle course navigation compared to existing methods may have affected the perceived workload scores, but these effects cannot be isolated because participants were surveyed after completing all experiments. Mobilio had a higher perceived mental demand but decreased effort, frustration, and physical demand compared to a white cane, suggesting that the additional information Mobilio provides the user through audio feedback compensates for the additional cognitive burden of interpreting this audio feedback. Open-ended participant survey responses generally reported that Mobilio provided intuitive audio feedback that was useful for staying on the right course, with the majority of users including one or more of these ideas: "It's simple, easy to use, and consistent." (Extended Data Table 8, question 1). The audio parameter personalization may have contributed to the intuitive audio feedback. All participants but one were interested in using Mobilio in the future (Extended Data Table 8, question 3). Mobilio's high usability and low workload scores suggest it may be feasible for everyday use.

Mobilio improves mobility metrics during navigation and may be feasible for everyday use, but additional work is needed to validate efficacy. Additional sensor and algorithmic testing across a range of navigation tasks, geographic locations, weather conditions, and during extended community use is necessary to assess and improve safety. At-home testing would require additional system development, such as a user interface to allow people with BVI to select their desired destination using a screen reader. Other designs for a more form-fitting chest strap, a waist harness, or lanyard around the neck could make smartphone placement easier and less conspicuous. Assessing alternative locations for wearing the smartphone on the body or alternative devices such as smart glasses paired with a phone is important to ensure the system meets users' needs for regular wear. A pilot test of a blindfolded participant without BVI enabled navigation of the community path when wearing the smartphone on a lanyard around the neck or in a clear waist belt, supporting that Mobilio can guide users when the smartphone is worn at different locations on the body.

This work advances the development of accessible and effective human-centered assistive technology by personalizing assistance for people with BVI. Mobilio's design as a smartphone application makes it a widely accessible tool for users, researchers, and clinicians in the BVI community. This smartphone-based approach to real-world personalization, computer vision, and sensor fusion could be extended to navigation for mobile robots or other assistive technology. The experimental results indicate that Mobilio may offer an effective solution to tackle major navigation challenges and improve metrics like heading error, navigation time, and the number of verbal navigation corrections needed compared to standards of care. Multi-modal sensing and sensor fusion algorithms played a key role in achieving the same reliability as human-level guidance during outdoor navigation. These performance improvements may help people with BVI increase their mobility, independence, and a variety of health outcomes. The unique needs of each participant resulted in widely varied personalized audio parameters across the group, suggesting personalization may be increasingly important for assistive technology to meet the diverse needs of heterogeneous patient populations. This study demonstrates the emerging potential of personalizing wearable assistive technology to enable essential community activity in many people's everyday lives.

## Methods

### Experimental design

The research objective was to evaluate the effects of personalized audio feedback and navigation guidance provided by Mobilio during walking. We used an iPhone 12 Pro with the Mobilio application for all experiments. We conducted two power analyses based on the first five participants and found that a sample size of eleven participants was necessary for the planned experiments. The first power analysis found a required sample size of ten participants based on the paired percentage differences in validation heading errors between generic and optimized parameters with a mean of 0.137 and a standard deviation of 0.133. The second power analysis found a required sample size of eleven participants based on the paired percentage differences in community path navigation time between Mobilio and Google Maps with a mean of 0.111 and a standard deviation of 0.108. Each of these power analyses used a power of $1 - \beta = 0.8$ and a significance level of $\alpha = 0.05$. We collected data from 14 participants, which was more than the minimum number determined by the power analysis, to provide a factor of safety in case data from any participants were found to be unusable during later analysis. All participants that were recruited completed the experiments, and all data from all participants were included in the analyses. All participants had at least 1 year of experience walking with a white cane, minimizing the effects of training that can occur with a white cane while performing walking tasks. All participants were volunteers and provided written informed consent before completing the protocol (IRB23-0585), which was approved by the Harvard University Institutional Review Board. Of the 112 people with BVI who completed the survey on navigation aids, 99 people completed the survey under the protocol IRB-55295, which was approved by the Stanford University Institutional Review Board. Consent was obtained for publication of identifiable images of research participants. Participants used an iPhone 12 Pro smartphone strapped to their chest with a chest mount, Shokz OpenRun open-ear headphones to receive spatial audio feedback, and a standard white cane. Participants used their own white cane. Each of the experiments is described in the following sections. We used repeated measures ANOVA to determine whether differences in outcomes across navigation assisted conditions were different from zero.

### Smartphone application design

Mobilio is a smartphone application created using Unity version 2023.2.13f1. Version 5.1.2 of the Unity ARFoundation package provides access to RGB camera images, depth images, and visual-inertial odometry to accurately track the smartphone's movement and orientation in a 3D world frame. Depth images rely on RGB camera images and use LIDAR if available. When the application is launched, the smartphone's position in the world frame is (0,0,0), the heading in the world frame is 0, and the vertical axis is aligned with gravity. The world frame heading increases as the user turns to the right. The application receives images and processes data at 30Hz. Versions 5.1.2 of the Apple ARKit XR and Google ARCore XR plugins enable deployment to iOS and Android devices, respectively. Unity Sentis version 1.6.0-pre1 is used to run the semantic segmentation model on the smartphone GPU. UniTask version 2.5.5 allows for the model layers to be split up and run over multiple frames to improve performance.

Mobilio's design as a smartphone application enables it to be complementary to existing solutions for people with BVI, including other navigation tools like white canes or guide dogs. Users may use different applications on the same device to handle additional tasks that Mobilio does not perform, such as reading text or using vision language models to describe environments. Future work may incorporate such tasks into Mobilio to encompass a larger set of navigation tasks without requiring users to switch between applications.

**Spatial audio feedback**

Mobilio provides continuous spatial audio feedback to minimize users' heading error. Audio feedback can effectively convey rich information with verbal and nonverbal cues and is commonly used in navigation aids for people with BVI[59]. Previous work used deep learning methods[60] to decrease over-rotating when performing turns with a turn-start and a turn-end instruction. The deep learning method decreased average angle over-rotation from 30.3° to 18.8° but was limited due to being trained on data from a small number of individuals that may not have adapted well to a diverse set of participants. Continuous audio feedback reduces rotation error compared to single-impulse sounds[61]. Participants using Mobilio had an average over-rotation angle of 16.2° during turn-and-walk validation experiments, indicating that the continuous audio feedback provided by Mobilio was similarly effective in preventing over-rotating compared to existing methods. However, continuous audio feedback may become irritating over time, can reduce a user's ability to hear other environmental sounds[62], and may be difficult to hear in noisy environments[63]. Future work could allow the system to be silent when the user is facing the correct direction, or potentially explore other methods of feedback such as smartphone-based haptic feedback, which BVI users may prefer in noisy outdoor environments but can only convey simple instructions.

Spatial audio feedback is provided with bone conduction headphones. Many previous ETAs that provided spatial audio used over-ear or in-ear headphones that blocked environmental sounds, which is not acceptable for BVI users[64]. We chose to use bone conduction headphones as they are less invasive than other headphones and do not block sounds from the environment. Some spatial audio approaches incorporate spatial elevation cues which may be useful for communicating the elevation of obstacles, but elevation cues are more difficult to accurately understand compared to horizontal cues, particularly for people with BVI[64]. Because of this, we chose a simplistic spatial audio approach that only provides spatial audio cues on the horizontal plane, which is adequate for communicating directions for users to go when navigating and avoiding obstacles.

To provide spatial audio feedback, a simulated audio source is placed in the Unity world frame 1 meter away from the user and in a spatial audio direction relative to the user. Audio feedback is parameterized with three parameters: audio rate ($r$), pitch range ($p$), and scaling factor ($s$). We personalize these parameters, which are initialized with a generic set of values: $r$ = 3Hz, $p$ = 0.5, $s$ = 1. The relative spatial audio direction is equal to the heading error multiplied by a fixed spatial scaling factor ($s$). For example, if the user should turn 40 degrees left and the scaling factor is 1.5, the audio will play as though it is 60 degrees to the user's left. The scaling factor can range between 0.5 and 2. Audio feedback consists of a repeated beeping sound that is 0.09s long. The rate at which beeps are played is defined by the audio rate parameter ($r$). The audio rate can range between 0.5Hz and 11Hz. The beeps play at different pitches based on the user's absolute heading error. The pitch range ($p$) defines how much the pitch can vary. The beep plays at a pitch of 1 when the user is facing the desired heading and has a heading error of 0. As the user turns away from the desired heading, their absolute heading error increases and the pitch decreases down to the minimum value of 1 - $p$. The pitch range can range between 0 and 0.8. Typically, decreasing the pitch of the beep also makes the beep take more time to play which reduces the maximum possible audio rate. To account for this, we decrease the pitch of the beep and then resample it to preserve the original length of 0.09s and have a uniform sampling frequency of 44100 Hz.

**Semantic segmentation model**

We trained a lightweight semantic segmentation model to enable Mobilio to provide accurate path guidance. There are few datasets for segmentation of street scenes that contain a large number of human-egocentric images[65]. Most segmentation models are trained on vehicle-egocentric datasets such as Cityscapes[66], making them less ideal for pedestrian use. Research into semantic segmentation on mobile devices is limited[48]. A few existing models trained on datasets with pedestrian images such as Mapillary Vistas[49] were too large to run well on a smartphone. To address this gap, we trained a lightweight segmentation model called PP-MobileSeg-Tiny[48] on the Mapillary Vistas dataset, which contains 18000 training images and 2000 validation images from both human and vehicle viewpoints with 66 visual object categories. We augmented the dataset to focus on identifying 8 surface categories that are important for pathfinding during walking: sidewalk, road, plain crosswalk, zebra crosswalk, curb, curb cut, terrain (such as grass), and covering (such as manhole or grating). The model was trained on images of size 480 by 480 pixels with a batch size of 4, providing 225,000 iterations for a total of 50 epochs. Training took approximately 22 hours with an NVIDIA GeForce RTX 4090. During training, dataset images were preprocessed by resizing the short side of the image to 480 pixels, taking a random 480 by 480 pixel crop, and applying a random horizontal flip with probability 0.5 and random uniform distortion factors between 0.5 and 1.5 to the brightness, contrast, and saturation. The RGB values for each pixel in each image, which range from 0 to 1, were normalized by subtracting (0.4217, 0.4606, 0.4720) and dividing by (0.2646, 0.2754, 0.3035). These values represent the mean and standard deviation RGB pixel values across all images in the Mapillary Vistas dataset. We used Lovasz-Softmax loss[67] to help deal with class imbalance and an AdamW optimizer[68] with a weight decay of 0.01. The learning rate began at 1e-6 and linearly increased to 0.0006 over the first 1500 iterations, then followed a polynomial rate decay[69] with a power of 1.

We deployed the trained model to run on the smartphone. The model was exported to the ONNX format using opset version 15 to make it compatible with Unity Sentis. The smartphone camera provided 640 by 480 RGB images which were center cropped to 480 by 480 and used as inputs to the model. We used Sentis to split up the layers of the model and execute them over multiple frames using the smartphone GPU. The model outputs were 120 by 120 images with each pixel being assigned one of the possible surface categories. Model inference was performed over multiple frames so that the model ran at 6Hz on the smartphone alongside the other sensing and navigation features of Mobilio.

We tested our model on the 2000 validation images in the Mapillary Vistas dataset. Each image in the validation set was preprocessed using the same steps as the training images. Although most dataset images were taken during the day, pilot testing of Mobilio at night indicates that semantic segmentation remains accurate enough for navigation in the presence of street lighting. The model contains approximately 1.5 million parameters and requires 627 million floating point operations (FLOPs) to perform inference on a 480 by 480 image. For comparison, we also tested a DeepLabv3+ model[70] with a ResNet-50 backbone[71] that was trained on the Mapillary Vistas dataset by OpenMMLab[72]. The DeepLabv3+ model achieved similar performance to our model (Extended Data Table 9) but contains approximately 41 million parameters and requires 156 billion FLOPs to perform inference on a 480 by 480 image. The model's large size and complexity means the smartphone cannot run the model at more than 0.2Hz which is not feasible for real-time use.

To account for reduced image quality and motion artifacts during walking, we evaluated our segmentation model on images with motion blur applied. Motion blur was simulated by applying a 2D convolution using a diagonal matrix kernel with values equal to $1/r$, where $r$ is the size of the kernel. The model has less than 10% reduction in the mean IoU score for motion blur kernel sizes of 5 or lower, with performance decreasing at a kernel size of 6 or larger (Extended Data Fig. 3). In practice, motion blur during walking appears visually similar to a kernel size of 3 (Supplementary Video 2), indicating that the model is suitable for use with motion artifacts during walking.

**Navigation algorithms**

We developed algorithms for obstacle detection and avoidance, GPS waypoint navigation, and path guidance that balance accuracy and computational complexity while running in real time on the smartphone. These algorithms fuse continuous streams of sensor data to determine a desired direction to guide users towards using spatial audio feedback. The algorithms run efficiently on the smartphone at 30Hz. The smartphone battery consumption was 64% over the course of a 115 minute experiment with one participant. The smartphone was active with Mobilio running for the full duration of the experiment. Based on this pilot test we expect approximately 3 hours of navigation with Mobilio on a single charge. The algorithms are detailed in the following paragraphs.

<u>Obstacle detection and avoidance</u>

Obstacle detection is performed by creating a 3D occupancy grid map using depth images and visual-inertial odometry. The occupancy grid incorporates information from all previous depth images to map the user's surroundings. This is important because the latest depth image on its own may not provide enough information for obstacle detection due to the smartphone camera's relatively restricted field of view. Additionally, an occupancy grid is computationally inexpensive to maintain and update, enabling real-time processing on the smartphone.

The environment is represented in the world frame as a 3D grid of cubic cells of side length 0.1 meters, where the goal is to determine whether each cell poses a collision risk to the user. Each grid cell is assigned a score indicating whether it should be treated as occupied. The occupancy grid is filled using 256x192 depth images. Depth images are initially received in the Unity XRCpuImage format. The data in each image is converted into a byte array. Different smartphones may receive the image data in different formats that can each be converted to one standard format. For example, an iPhone 12 Pro receives 4 bytes of information per pixel in the depth image which must be converted to a 32-bit floating point value describing the depth in meters. An Android phone might receive 2 bytes of information per pixel which must be converted to a 16-bit unsigned integer describing the depth in millimeters. We determine the method of conversion based on the number of bytes per pixel which is given by the XRCpuImage struct. We can similarly receive confidence values for each pixel in the depth image that represent the confidence in the accuracy of the estimated depth values. Each confidence value is an 8-bit unsigned integer. On iOS devices, the confidence values are either 0, 1, or 2; on Android devices, the confidence values are integers ranging from 0 to 255, inclusive. For each depth image, we iterate over each pixel in the image and extract the corresponding depth value. If the depth value for a pixel is less than 4 meters and the confidence value is strictly greater than half of the maximum possible value, it is projected to a position relative to the camera using the depth value, the pixel coordinates, and the camera's principal point and focal length. We only use points that are within 4 meters of the smartphone and have a high confidence value to limit memory usage and maintain accuracy. The point in the local camera frame is projected to the world frame using a 4x4 transformation matrix that is calculated from the current position and orientation of the camera in the world frame. The occupancy grid cell containing this point then has its score incremented by 1. We limit the maximum possible memory usage of the occupancy grid by deleting grid cells that are more than 5 meters away from the smartphone. To enable the occupancy grid to accurately track moving objects such as a pedestrian walking by, the scores for grid cells are exponentially decayed by a factor of 0.9 while they are in view of the camera. We determine whether a grid cell is visible to the camera by checking whether its center is on the positive side of each of the top, right, bottom, and left camera planes. The camera planes represent the limit of what the camera can see. The planes are continuously updated as the user moves by using points projected from the top-right, bottom-right, top-left, and bottom-left pixels in the depth image. Each plane is defined using three points. For example, the top plane is defined using the camera position and the projected top-right and top-left points. This approach maintains user safety by preserving the mapping of objects outside of the field of view of the smartphone camera, while ensuring that moving objects are also properly updated.

After processing a depth image and completing all updates to the occupancy grid, we use the occupancy grid to estimate the location of the ground, detect potential collisions with obstacles, and direct the user to avoid collisions. We iterate over all cells in the occupancy grid. If the cell score is greater than a threshold it is considered occupied. We estimate the vertical location of the ground that the user is standing on by computing the average vertical location of all cells that are occupied, below the smartphone, and within 0.3 meters of the smartphone in the horizontal direction. A cell is treated as an imminent collision if it is occupied and in front of the user. The position of each occupied cell in the world frame is transformed to the local camera frame to make it easier to check if it is in front of the user. A cell is in front of the user if it is within 1.5 meters in the forward direction from the smartphone, within 0.2 meters to the right or left, and above the ground and below the height of the user.

When there is an imminent collision between the user and an obstacle, we determine the best direction for the user to move towards to avoid any obstacles. All cells in the occupancy grid that are occupied, above the ground, and below the height of the user are aggregated and used to populate a horizontal 2D search grid that has the same cell length and covers the same horizontal area as the 3D occupancy grid map. Each cell in the search grid is assigned a value of 0 if the user can move over it without colliding with an obstacle, and is otherwise assigned a value of 1. The populated search grid is used to search for the best line direction from the user. We repeatedly cast lines from the user's position on the search grid at different angles relative to the horizontal direction that the user is currently facing. The first line's direction relative to the user's direction is 0 degrees, then 2, -2, 4, -4, and so on until 90 and -90 degrees. We use a fast voxel traversal algorithm[73] to determine the cells in the search grid that each line intersects with. If a line travels for 2 meters without intersecting with a cell that has a value of 1, we use the angle of that line as the direction for the user to turn to avoid an obstacle. Otherwise, the line that traveled the furthest distance is used. This method may be augmented to take into account the width of the user by iterating over each of the search grid cells with a value of 1 and setting nearby cells to also have a value of 1. However, we found that this caused issues when using our simple line direction method in narrow spaces. The use of a 2D search grid enables more complex algorithms such as the A* algorithm[74] that could be used in place of our method.

<u>Comparison between depth estimation with LIDAR or monocular depth estimation</u>

LIDAR sensors are often used to accurately measure distances to the environment, but not all smartphones have a LIDAR sensor. Some monocular depth estimation models can use RGB images to estimate metric depth, or real-world distances to environments. However, these models have difficulty generalizing to images with different scales such as indoor and outdoor images. We implemented a method that uses the depth of AR feature points to estimate scale and shift coefficients for a monocular metric depth model in real time on a smartphone. Each monocular depth value is scaled by the scale coefficient and then shifted by the shift coefficient to estimate the true real-world distance to the environment. Feature points are points that are tracked in the world frame by ARKit or ARCore which provide depth values that are more accurate than other points in the camera view that can be used to calibrate monocular depth values. Given a set of feature points $p$, each point is projected to a pixel coordinate in the smartphone screen space to get a corresponding monocular depth value. The scale and shift values are computed using 200 iterations of random sample consensus (RANSAC) algorithm to determine values that best align the feature point depths with the monocular depths with the fewest number of outliers. A feature point is considered an outlier if its scaled and shifted monocular depth has a percentage error of at least 0.05 compared to its world frame depth. The percentage error between a ground truth distance value $d$ and a distance estimate $\hat{d}$ is equal to $\left|\widehat{d-d}\right|/d$. If the number of inliers is at least 40 and the ratio of the number of inliers to the total number of points is at least 0.7, we perform a linear regression on all inliers to calculate the final scale and shift coefficients. This heuristic ensures that there are sufficient points to estimate the scale and shift with sufficient confidence approximately 50% of the time during outdoor navigation.

We used a monocular depth estimation model called Depth Anything V2[51] (DAV2) to evaluate the feasibility of performing real-time depth estimation from RGB images on a smartphone. The model was a pre-trained Depth Anything V2 model with a ViT-S encoder[75] that was fine-tuned on the KITTI autonomous driving dataset[76] to perform outdoor metric depth estimation. We exported the model to the ONNX format using opset 17 to deploy it on the smartphone. The model contains approximately 25 million parameters and can run at 3Hz on the tested smartphone with an input and output size of 238x238 without impacting the other sensing and navigation features of Mobilio.

We evaluated the depth estimation accuracy of LIDAR and DAV2 by having a participant wearing Mobilio walk in a motion capture lab space towards five representative obstacles: a stop sign, a person, a tree branch hanging at head height, a fire hydrant, and a 2cm wide pole. We attached motion capture markers to the smartphone and to each object to obtain ground-truth distances between the smartphone and the objects. Each obstacle was approached five times, with each trial starting from 5 meters away. We compared the motion capture distances with distances extracted from LIDAR measurements and the scaled and shifted DAV2 outputs (Extended Data Fig. 8). LIDAR and DAV2 had average percentage errors of 5.7% and 14.6% and were able to accurately estimate distances to objects, especially at distances of less than 3 meters. DAV2 was able to accurately estimate distances due to the consistent distribution of the scale and shift coefficients based on distances to the obstacle (Extended Data Fig. 9). Scale decreased linearly as the distance to the obstacle decreased, likely due to the back wall behind the obstacle. Shift was approximately 1 regardless of distance. These results indicate that monocular depth estimation may provide sufficiently accurate depth estimation to perform obstacle detection on smartphones without LIDAR.

GPS waypoint navigation

Navigation along GPS waypoints relies on the ability to calculate the bearing and distance from one GPS waypoint to another. For example, when a user approaches a turn, we calculate distance to the turn and the angle of the turn. These are needed to determine when the user has reached the turn and how to direct to the user. Suppose the user is moving from waypoint A to waypoint B, where they then will turn and walk towards waypoint C. The turn angle is determined by calculating the bearing from A and B and subtracting it from the bearing from B to C. Calculating the bearing between two GPS coordinates can be complex due to the Earth's curvature. Let $(\varphi_1, \lambda_1)$ and $(\varphi_2, \lambda_2)$ be the latitude and longitude of two points in radians, with $\Delta\varphi = \varphi_2 - \varphi_1$ and $\Delta\lambda = \lambda_2 - \lambda_1$. We use a simple formula that accurately estimates the bearing between two nearby points: $atan2(\Delta\lambda \cdot \cos\varphi_2, \Delta\varphi)$. This formula ignores curvature but remains accurate for any realistically close pair of GPS coordinates that are used for navigation during walking. For example, if the two points are within one kilometer of each other, this formula will have an error of less than a hundredth of a degree. The great-circle distance between the two points may be calculated using the Haversine formula[77], but similar to estimating the bearing, we use a simple formula that accurately estimates the distance in meters between two nearby points: $\sqrt{(\Delta\varphi \cdot C)^2 + (\Delta\lambda \cdot C \cdot \cos\varphi_2)^2}$. In this formula, φ and λ are in radians, and $C$ is a constant equal to $111320 \cdot \frac{180}{\pi}$ meters. The constant $C$ converts latitude values to meters while $C \cdot \cos\varphi_2$ converts longitude values to meters. For any two points within one kilometer of each other, this formula will have an error of less than a tenth of a meter.

Path guidance

Mobilio provides path guidance to complement turn-by-turn directions which may be inaccurate when relying only on smartphone GPS and compass measurements. During turn-by-turn navigation, the user navigates along a sequence of GPS waypoints. We can calculate the world frame direction towards the current target GPS waypoint using the user's world frame heading, compass heading, and the bearing towards the target waypoint. However, this direction may be inaccurate due to errors in GPS or compass measurements.

Path guidance along walkable paths is provided by using outputs from our semantic segmentation model to create a horizontal 2D surface grid map of nearby street surfaces. The goal is to identify areas and paths where the user may safely walk. For example, users can walk on sidewalks and crosswalks, but not on roads. The surface grid consists of square cells of side length 0.2 meters. Each cell is assigned a true or false value indicating whether the user can walk over it. The vertical level of the surface grid in the world frame is set to the ground level that was determined from the 3D occupancy grid map. When the segmentation model is not currently running, the next RGB camera image received by the application is used as an input to the segmentation model. The model outputs an image where each pixel corresponds to one of the segmentation categories. When the model begins running, the position and orientation of the smartphone camera in the world frame is recorded so that it can be used later while processing the model output. We update the surface grid by iterating over all cells in the surface grid that are within 15 meters of the user, projecting the position of the center of the cell in the world frame to a pixel coordinate in the smartphone screen space, and converting that to a pixel coordinate in the model output image. The category of the pixel in the output image is used to assign the original cell a value. If the cell is behind the user, or if its projection to the output image does not fall within the view of the image, it is not updated. This process ensures that all visible grid cells are updated with the semantic segmentation model output.

The updated surface grid is used to search for potential paths for the user to take. The current direction towards the target waypoint is used to search for potential paths on the surface grid. Similar to the algorithm described in ‘Obstacle detection and avoidance’, we cast lines from the user’s position on the surface grid. The lines are cast at 1 degree intervals between -90 and 90 degrees relative to the target direction. Lines that travel at least 5 meters before intersecting with a non-walkable cell are considered to be valid and part of a potential path. We iterate over the valid lines and divide them into subsets, with divisions occurring where the valid line angles are not consecutive. For example, given a set of valid lines cast at 10, 11, 12, 15, 16 degrees relative to the target direction, the first three lines comprise one subset and the last two comprise a second subset. Each subset represents a potential path. The angle of the path is determined by the angle of the longest line within the subset. The subset with the closest angle to the target direction is used as the direction for path guidance if it is within 30 degrees of the target direction, otherwise the target direction is used.

Degraded paint in crosswalks or sidewalks can cause the segmentation model to misclassify areas of crosswalks or sidewalks as non-walkable, leading to gaps in the surface grid (Extended Data Fig. 2b). When performing a line cast to search for walkable paths, we address minor gaps by continuing to search up to 3 grid cells ahead in the surface grid even when encountering non-walkable cells. However, if the sidewalk or crosswalk is in major disrepair, no suitable path will be found on the surface grid and Mobilio will revert to relying only on GPS for navigation.

Smartphone GPS measurements are corrected based on knowledge of the path and upcoming turns. Smartphone GPS measurements may have errors of greater than 5 meters and these errors may change over time. To account for these changes, we maintain latitude and longitude offset values that are added to the raw GPS measurements to provide an estimate of the user’s true GPS position. We adjust the latitude and longitude offsets to place the user’s location on the expected path. The user’s current GPS position is projected onto the line that contains the previous waypoint and the current target waypoint. If the projection distance is less than 12 meters, we assume that the user is on the correct path and adjust the offsets such that the user’s position with the offsets added is on the path. This corrects the user's GPS position sideways along the path, but not forward or backward to avoid causing issues when determining when the user has reached the target waypoint.

Existing smartphone-based turn-by-turn navigation systems such as Google Maps or the Wayfinding and Backtracking apps[78] provide turning prompts before the user reaches the turn to account for possible localization errors from GPS or inertial navigation. For example, Google Maps may provide turning instructions up to 30 meters in advance of a turn. For Mobilio, access to camera data enables more accurate turning prompts by using VIO to track user positions with minimal error and semantic segmentation to detect turns. A previous smartphone-based approach to intersection detection from Corridor-Walker[59] used LIDAR to construct a 2D occupancy grid map of an indoor corridor and applied a YOLOv3 model to the map to detect turns. However, this approach may not work in outdoor environments without walls.

We implemented an algorithm to detect upcoming turns that a user may miss when relying only on GPS measurements. After determining the direction of the path, we search along the path for the next turn. The turn angle is the world frame direction between the current target waypoint and the next target waypoint. A line is cast on the surface grid in the direction of the path using the same traversal method described earlier. Starting at 2 meters along the path line and every 0.1 meters after, we cast a line branching from the path line in the direction of the turn angle that stops upon intersecting with a non-walkable cell. This branch is longer when on a turn. We also cast an opposing branch in the opposite direction of the turn angle to estimate the full width of the path. We record the main branch length as well as the total length, which is the sum of the main branch length and opposing branch length. These lengths are longer when on a turn which enables turn detection. After all main branch lengths and corresponding total lengths are recorded, we iterate over them to determine which branches may represent a turn. A main branch is considered to be part of a turn if it is at least 3.5 meters longer than the median of all recorded main branch lengths, and the corresponding total length is at least 3.5 meters longer than the median total length. The valid branches are divided into subsets based on the branch starting points along the original path line, with divisions occurring between non-consecutive valid branches. The process of dividing branches into subsets is similar to the one used when searching for paths on the surface grid. Each subset of branches represents a potential turn. The distance of the turn from the user is determined by the average location of the starting points of the branches in the subset. For each subset we determine the expected location of the turn based on the subset branch distance from the turn, the angle of the path line, and the user's current estimated GPS position. If the expected location is within 8 meters of the target waypoint, we set the latitude and longitude offset values such that the expected turn location with the offset is identical to the target waypoint.

While we modify GPS measurements to correct errors based on knowledge of the path and upcoming turns, we do not modify world frame positions or orientations such as in RouteNav[79]. VIO tracks positions and orientations with 1.2 meters of drift over a 300 meter path, or 0.4% error and is reliable to use for local path guidance. Additionally, RouteNav relies on pre-defined external knowledge of walls, no-go zones, and GPS-denied zones to better account for drift.

We observed that smartphone compass measurements were often inaccurate when walking outdoors due to interference from nearby metals, electronics, and other environmental factors. We solve this problem by estimating an offset between the phone's world frame heading and the true compass heading. This approach provides compass headings within a few degrees of error. It depends only on the accuracy of the visual-inertial odometry performed on the smartphone, which we observed to provide more consistent heading estimates than the compass. The heading offset is estimated by calculating the best fit rotation matrix between a set of GPS coordinates and positions in the world frame. Mobilio receives GPS coordinates at a rate of 1Hz. Each GPS coordinate is saved along with the smartphone's horizontal position which excludes the vertical component. Data saving is paused until the user moves at least 0.3 meters horizontally from their last saved position. When twenty pairs of coordinates have been saved, we calculate a rotation matrix that best aligns the coordinates. Let $P$ be a 20 by 2 matrix containing twenty world frame positions and let $Q$ be a 20 by 2 matrix containing the corresponding GPS coordinates. The matrices $P$ and $Q$ are transformed by subtracting the mean coordinate of each matrix. The first column in $Q$ is scaled by $C$ and the second column is scaled by $C \cdot cos\ \varphi$ to convert $Q$ to meters. These variables are defined in 'GPS waypoint navigation'. The covariance matrix $H = P^{\top}Q$ and the singular value decomposition $H = U\Sigma V^{\top}$ are calculated. The rotation matrix that aligns $P$ with $Q$ is $R = VU^{\top}$. If the determinant of $R$ is negative then it also performs a reflection. The reflection is removed by negating the values in the second column of $V$ and recalculating $R$. The angle of rotation is $atan2(R_{2,1}, R_{1,1})$, where $R_{2,1}$ refers to the value in the second row and first column of $R$. This angle is added to the user's heading in the world frame to estimate the user's compass heading.

While we align VIO position trajectories with GPS position trajectories to calculate a rotational offset, we do not apply the same process to calculate a translational offset between VIO and GPS, as it may be unreliable. As an example, if a user begins navigation by walking north in a straight line next to a tall building, GPS may consistently measure the user to be 3 meters west of their actual position. In this example, we would be unable to rectify the GPS error until the user takes a turn. Therefore, we rely on detection of turn locations in the world frame, for which the corresponding GPS location is known, to align VIO with GPS.

**Participant experiments**

Audio feedback learning

The training time required for users to adapt to using assistive technology can vary depending on the complexity of the device. We performed an experiment to identify the training time needed for participants to learn to use the audio feedback from Mobilio to reduce their heading error. This study was performed before the experiments with BVI participants to validate that personalizing audio feedback parameters would help participants become expert users and identify how long of a personalization experiment was required. Ten participants without BVI (n = 10; 5 men and 5 women; age, 23.9 ± 1.51 yr; body mass, 69.2 ± 9.67 kg; height, 1.73 ± 0.06 m) were blindfolded and instructed to turn in place to face a series of randomized desired headings provided by audio feedback. The experiment consisted of repeating blocks of training trials and evaluation trials. Each trial tested three different target headings. During the training trials, audio feedback was personalized to each participant with CMA, using the same personalization approach described in this study for participants with BVI. During evaluation trials, participants received a pre-defined set of audio parameters to evaluate how well they learned to use the audio feedback system. This learning was modeled by fitting an exponential curve to the normalized heading errors during evaluation for all participants throughout the training process. The normalized average heading error for a trial is equal to the average heading error over the duration of the trial divided by the heading error at the start of the trial. Participants reached 86% of this steady-state exponential asymptote at 29 minutes, 90% at 34 minutes, and 95% at 44 minutes of training (Extended Data Fig. 10).

Personalizing audio feedback

The CMA-ES optimizer maintained an estimate of the best audio feedback parameters for the user, as well as a covariance matrix and a convergence parameter that dictated the shape and spread of the parameter search area. During optimization, the optimizer selected a generation of parameter sets to evaluate. Each generation consisted of seven parameter sets to provide the minimum number of samples to update the estimate of the best parameters during optimization[54]. Each parameter set was evaluated based on the user's normalized average heading error over the first five seconds after they encountered a turn. Average heading error was normalized by the heading error at the start of the turn so the magnitude of each turn didn't affect comparisons between parameter sets. Pilot tests indicated that turns normalized by magnitude could be reliably compared because participants had a similar rate of completing the turns regardless of the initial angle (Extended Data Fig. 4).

After each generation, the optimizer used a weighted average of the best performing parameter sets in the generation to update the estimate of the best parameters, the covariance matrix, and the convergence parameter. These updated estimates and matrices were sampled to select a new generation of parameter sets to evaluate. Audio feedback was experimentally personalized to each participant while turning and walking in an unobstructed 10 meter radius area in an open field. The area was centered around the participant's starting point. A new point was selected every 5 to 8 seconds. The point was chosen uniformly at random from the area that was within 4 to 8 meters of the participant, between 60 to 120 degrees to the right or left of the participant's current heading, and within the 10 meter area. Audio feedback guided the participant to turn and walk towards each point. The participant performed 196 turns totaling 28 generations of optimization over approximately 21 minutes of walking. The participant experienced one audio parameter set during each turn. Each parameter set was scored by the normalized average heading error over the first 5 seconds. The CMA optimizer's estimate of the best audio parameters for the participant after the 28 generations were considered the optimized parameters for that participant. A validation experiment compared the participant's performance with generic and optimized audio parameters. The participant completed turns for a set of 16 angles (60, 64, 68, …, 120 degrees) to the right or left for both the generic and optimized parameters. The 16 turns with the generic parameters and the 16 turns with the optimized parameters were randomized and presented in a double reversal order, totaling 64 turns over approximately 7 minutes of walking. We accounted for participant reaction time and maximum turning speed by subtracting the participant's lowest valid error across their trials from each of their heading error values. This more accurately captured the improvement towards optimal performance. We consider the error from a trial to be valid if the participant was not turning for the last second before the trial.

Community path navigation

We conducted experiments in a community setting to evaluate the path guidance provided by Mobilio, Google Maps, and a human guide. Participants walked along a set of predetermined GPS waypoints to navigate a 150 meter community path. Participants using Mobilio received spatial audio feedback that guided them along the path towards each waypoint. Mobilio provided text-to-speech voice directions on how to turn upon reaching a waypoint. This was not strictly necessary but minimized confusion that may occur from sudden changes in the spatial audio feedback. Mobilio and Google Maps both provided the same voice directions on the direction to turn to isolate the performance effect of the continuous audio feedback that Mobilio provided in addition to the voice directions. The user was provided with an audio cue to "take a slight right turn" when the turn angle was between 15 degrees and 45 degrees, "turn right" for turn angles between 45 degrees and 120 degrees, and "take a sharp right turn" for turn angles greater than 120 degrees. The turn angles were similarly partitioned for left turns. During the human guide condition, a research team member acted as a human guide and told the participants to turn exactly when they reached the turn. The turn instructions were identical to those provided by Mobilio and the same type of turn-by-turn instructions from Google Maps, with users being told to either "turn right" or "take a sharp right turn" for right turns. The human guide gave multiple turn instructions if the participant did not complete the turn on after the first instruction; these did not count as verbal corrections. Google Maps provides turn-by-turn voice directions during navigation and has a "Detailed voice guidance" feature[80] for users with BVI that provides more detailed directions. To eliminate any difference in performance due to pathing or waypoint differences between Mobilio and Google Maps, we developed a proxy Google Maps that relied on the exact same selected waypoints as Mobilio. While this reduces the generality of our testing to this specific path, it isolates the difference in performance between these two systems to the feedback that they provide, rather than differences in pathing or sensing. The proxy Google Maps used the raw smartphone GPS measurements and provided text-to-speech voice directions that closely simulated the directions given with the "Detailed voice guidance" feature active. Participants using the simulated Google Maps received an audio cue when navigation began that told them cardinal direction and distance to the next waypoint. For example, if the next waypoint was 40 feet northwest from the user, the audio cue told the participant to "head northwest for 40 feet". When approaching a waypoint, participants received an audio cue that told them the direction to turn and the distance to the next target waypoint. For example, if the participant needed to turn right and the next target waypoint would be in 35 feet, the audio cue would tell the participant to "turn right, then continue for 35 feet." The initial cardinal direction given when starting navigation may not be helpful for a user with BVI, but we aimed to simulate Google Maps as closely as possible, and participants always began navigation facing the correct direction which mitigated any confusion.

Participants walked back and forth between the two endpoints of the community path. The conditions were randomized with double reversal, as ABCCBA, for a total of 6 runs. We recorded the completion time and number of verbal navigation corrections provided for each run. If a user realizes that they are walking in the wrong direction even without verbal corrections, we expect that they may be able to correct themselves and complete the path given enough time, especially given the simplicity of the path. However, verbal corrections were provided when users walked the wrong way for 10 seconds or when they asked for assistance to avoid prolonging participant confusion. The most common source of errors requiring verbal corrections when using Google Maps was confusion around the early and imprecise turn directions, which was compounded by a lack of continuous audio feedback. Google Maps provided turn directions approximately 12 meters before users reached each turn and varied based on GPS error. Without continuous feedback, participants may not know exactly when to turn and Google Maps would introduce navigation errors where participants would walk past a turn or take a wrong turn which would require the research team to provide a verbal correction to help the participant reorient themselves and continue navigating on the path (Extended Data Fig. 11a). Participants that used shorelining to search for turns would miss the turns less often but would slow down when searching for a turn. These participants could still become confused when trying to turn too early, causing them to go the wrong way to search for a turn (Extended Data Fig. 11b). Some participants stopped and asked for clarification which was counted as a verbal correction. Continuous audio feedback from Mobilio may have helped participants turn the appropriate amount and at the appropriate time and helped them correct their directions within a ten-second period before requiring intervention from researchers (Extended Data Fig. 11a). To account for the reduction in completion time caused by the corrections provided by the research team during statistical analysis, the completion time for a run was increased by 30 seconds for each correction provided during the run.

Obstacle course navigation

We created a cardboard hallway that was 1.8m wide and 4.2m long. Two 0.85m wide obstacles were placed randomly at one of the 0.6m placement intervals along the walls, at least 1.8m apart, with one obstacle on each side of the hallway. The hallway obstacle course was not necessarily a realistic environment for outdoor navigation, but it was a suitable setting for evaluating the obstacle avoidance directions provided with spatial audio feedback from Mobilio. Participants navigated the hallway in two conditions, using either a white cane or Mobilio and a white cane. In each condition, participants navigated the hallway 10 times. The conditions were randomized with double reversal, as ABBA. We randomly selected 10 different obstacle placements to use in every condition. The ordering of the 10 obstacle placements was randomized for each of the first two conditions. For the third and fourth conditions, the obstacle placements were presented in the reverse order of the placements in the first two conditions. For each run, we recorded the time needed to reach the end of the hallway and the number of times the cane or the participant's body contacted the environment.

Comparison with previous methods

Most existing ETAs have not been experimentally evaluated with participants with BVI to determine whether they improve navigation in comparison to existing methods such as a white cane[5,27]. Here we review some of the ETAs that have been evaluated. The Augmented Cane[20] improved walking speed by 18% across indoor and outdoor navigation while reducing contacts in a hallway obstacle course, but the outdoor path was approximately 4 meters wide and had no sharp turns, which may have made it easier to follow than typical sidewalks. Another electronic cane decreased walking speed by 35% during indoor obstacle course navigation[81]. Corridor-Walker[59] significantly increased task completion time while decreasing contacts across a variety of routes. A system that used distance sensors worn on the body[22] increased duration while decreasing contacts during hallway tests. A system that used a depth camera mounted on glasses and provided audio and haptic feedback decreased contacts and navigation time during indoor obstacle course navigation after participants were trained to use the system in virtual environments[82]. Two commercial devices that were tested in an indoor obstacle course reduced collisions and reduced walking speed[83]. We note that cane contacts are not inherently negative; cane contacts were recorded to better understand the effect of obstacle avoidance from Mobilio on how participants use their canes. It is important to perform navigation tasks in representative community settings to evaluate ETAs in scenarios similar to how they will be used in everyday life.

Compared to obstacle avoidance devices, there are relatively few outdoor navigation systems for people with BVI. Two such systems are CaBot[23] or Drishti[16], although neither of these evaluated navigation time or walking speed.

**Methods references**

**Acknowledgements:** We thank A. Chin, C. Campillo-Rodriguez, A. Saffarini, and A. Yu for assistance with data collection; N. Fidalgo and P. Vegesna for assistance with pilot experiments; K. Baker, K. Garcia, A. Gracy, and the Carroll Center for the Blind for coordinating with participants; S. Slade and H. Liu for editorial suggestions.

**Funding:** National Science Foundation Graduate Research Fellowship DGE-2140743, Harvard Grid Accelerator, Amazon Greater Boston Tech Initiative, Harvard Dean's Competitive Fund for Promising Scholarship, and the Kempner Institute.

**Author contributions:** R.L. contributed to conceptualization, methodology, investigation, visualization and writing. P.S. contributed to conceptualization, methodology, visualization, and writing.

**Competing interests:** The authors declare no competing interests.

**Data availability:** All study data necessary to replicate this work are available in a public repository: https://github.com/Harvard-Slade-Lab/Mobilio. This includes all the experimental data from 14 participants that completed the study.

**Code availability:** Navigation algorithm code samples are available in a public repository: https://github.com/Harvard-Slade-Lab/Mobilio. This includes algorithms to perform obstacle detection and avoidance, GPS waypoint navigation, and path guidance.

**Supplementary Video 1**
This video shows a participant performing the three experiments in this study: personalizing audio feedback, navigating a community path, and navigating an obstacle course. Approximately two minutes of the audio feedback personalization experiment is played back 4 times faster than the actual speed to illustrate how the participant walked in a field and performed a series of turns to identify optimal audio parameters to meet their unique sensory needs. One participant is shown navigating the community path 4 times faster than the actual speed. One participant is shown navigating an obstacle course, first with only a white cane, and then with a white cane and audio feedback from Mobilio.

**Supplementary Video 2**
This video shows the navigation algorithms and spatial audio feedback the smartphone performs during both the community path and obstacle course. We recommend listening with headphones to experience the same spatial audio feedback as the participants. The community path navigation visualizes how the semantic segmentation model separates the path from other parts of the camera image to determine what spatial audio is needed to provide path guidance for the participant. The raw camera feed is shown on the left of the screen and the semantic segmentation model outputs processed in real time on the phone is shown on the right. The red arrow is a visual representation of the direction the user should turn to stay on the path, which is also provided as the desired heading with the spatial audio feedback. The community path navigation is played at 4 times the actual speed after the first two turns are completed. The obstacle course navigation visualizes the depth images on the right side of the screen, with a red arrow indicating the direction the obstacle avoidance model indicates the desired heading the user should turn towards to avoid a collision. The desired heading is provided with spatial audio.

**Supplementary Video 3**
This video includes representative background noise of the community path used for the navigation experiments. This community path is in an area where people are walking or sitting and talking, wind is moving the trees, construction can be heard on nearby buildings, and cars are traveling on neighboring roads.

**Extended Data Fig. 1: Processing sensor data to perform navigation capabilities. a,** Path guidance is provided by performing semantic segmentation on RGB camera images. Segmentation outputs are fused with visual-inertial odometry to create a 2D surface grid map. The surface grid is used to determine the direction of the path and correct errors in GPS position measurements. **b,** Obstacle detection and avoidance is performed using depth images that are derived from RGB camera images and LIDAR data, if available. Depth images are fused with visual-inertial odometry to create a 3D occupancy grid map. The occupancy grid is used to detect obstacles and provide directions to avoid collisions.

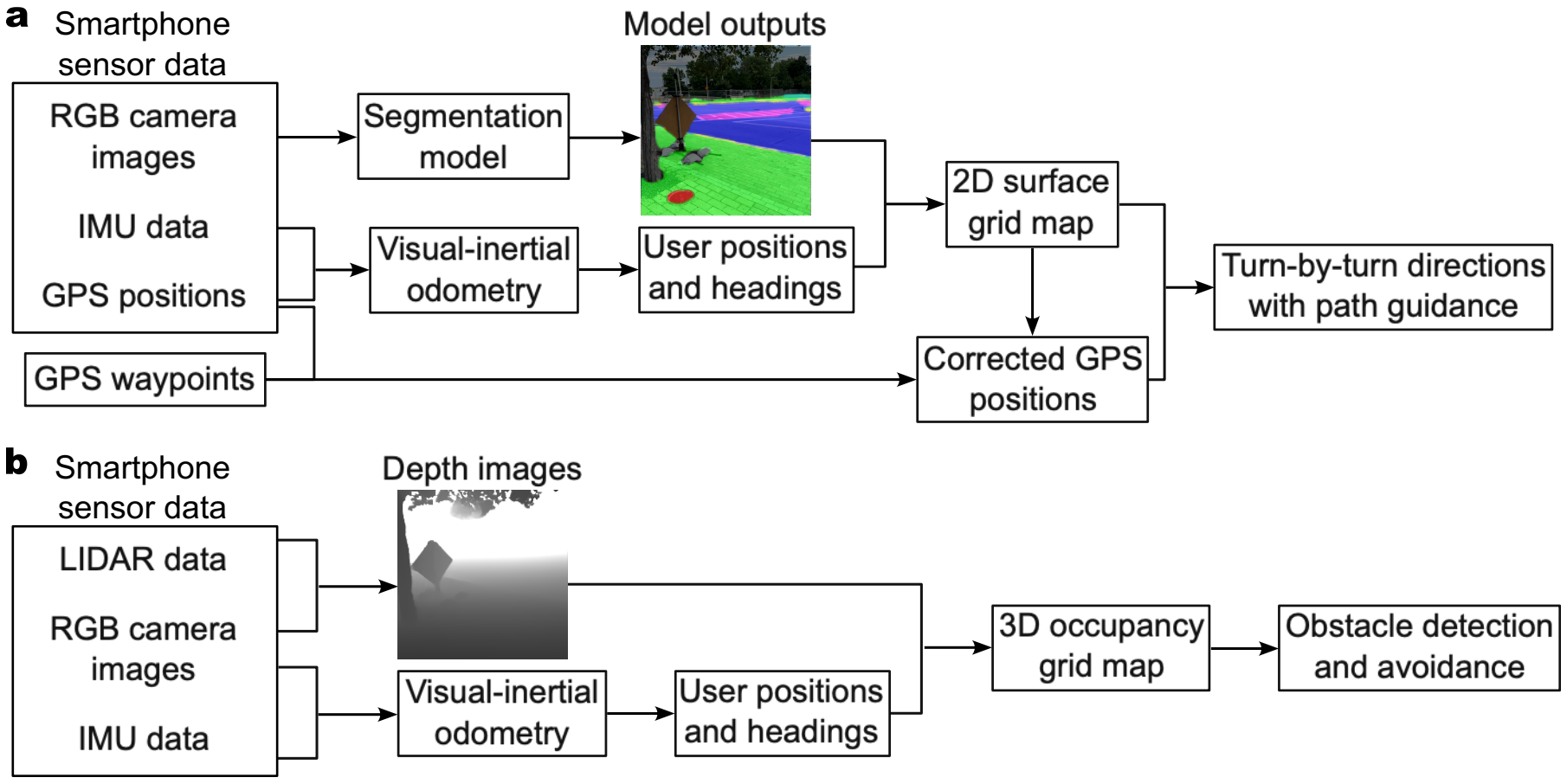

**Extended Data Fig. 2: Representative images and hand-selected challenging cases for the semantic segmentation model. a,** Ten representative examples of segmentation model outputs show segmentation across a variety of environments. The model can perform well in rainy conditions, on partially degraded crosswalks, and at night with sufficient lighting. The model is trained to identify surfaces including sidewalks, plain and zebra crosswalks, roads, terrain, coverings, and curbs. The model excludes non-surfaces such as pedestrians, vehicles, and poles. **b,** Ten qualitative examples of challenging segmentation cases. Sidewalks that are perceptually similar to roads may be misclassified. Walkable non-sidewalk paths may be classified as roads. Degraded crosswalks are more difficult to detect due to limited paint and less structured shape on the street. Inclement weather conditions such as snow can occlude surfaces, making them difficult to differentiate.

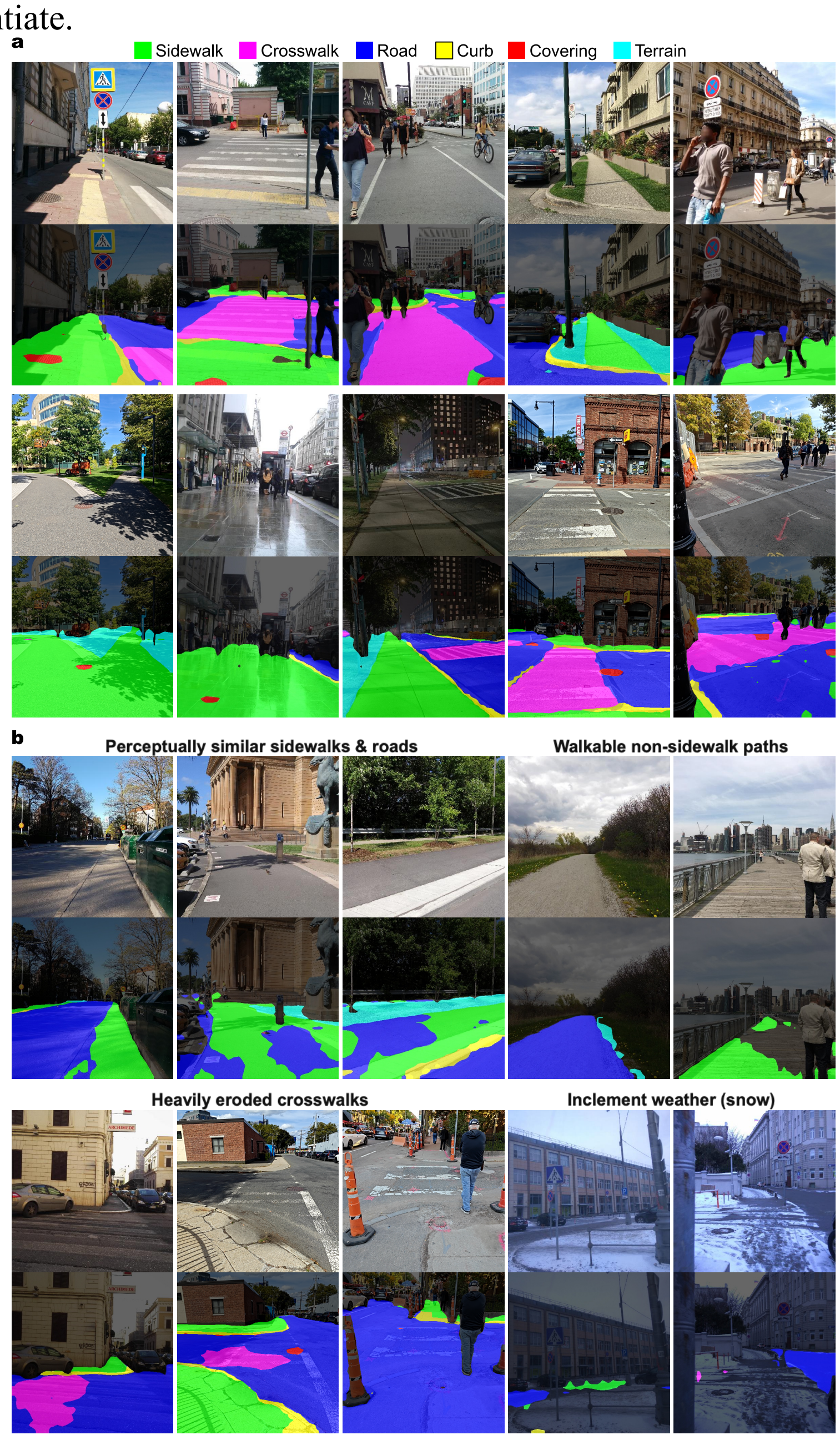

**Extended Data Fig. 3: Semantic segmentation model performance on blurred images. a,** Motion blur was simulated by applying a 2D convolution with a diagonal matrix. Model performance decreases at motion blur kernel sizes of 6 or greater. In practice, motion blur during walking appears visually similar to a kernel size of 3. **b,** Segmentation model performance was evaluated on the Mapillary Vistas validation set with increasing levels of motion blur applied to the images. The model has less than 10% reduction in the mean IoU score for motion blur kernel sizes of 5 or lower, with performance decreasing at a kernel size of 6 or larger.

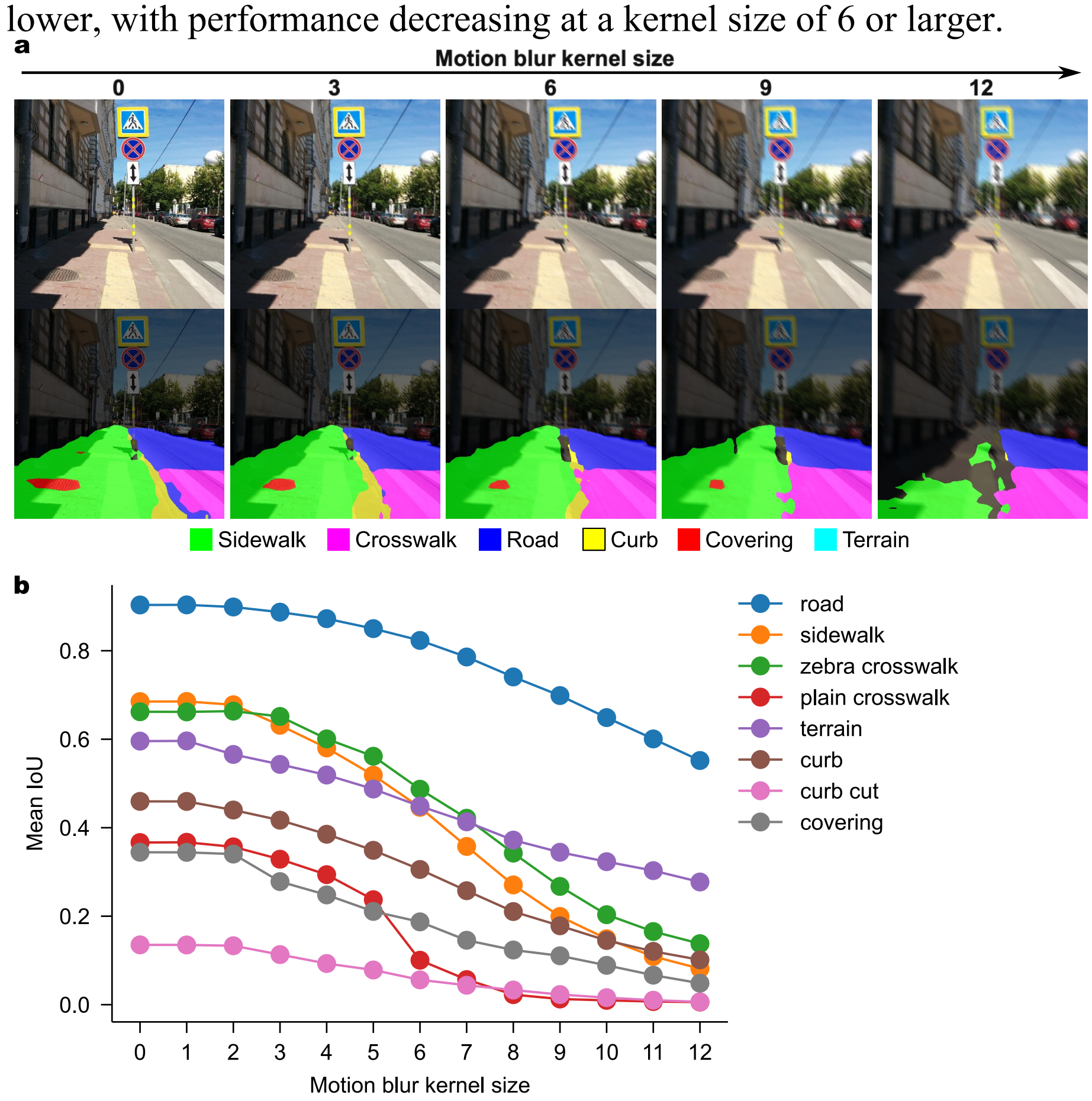

**Extended Data Fig. 4: Comparing normalized average heading error with different initial angles.** Pilot testing of spatial audio feedback indicated that turns of different magnitudes could be reliably compared by normalizing by the initial magnitude of the turn. One participant performed a set of turns with predefined angles (30, 40, 50, 60, 70, 80, 90 degrees). The ordering of turns was randomized. Each turn was performed 42 times.

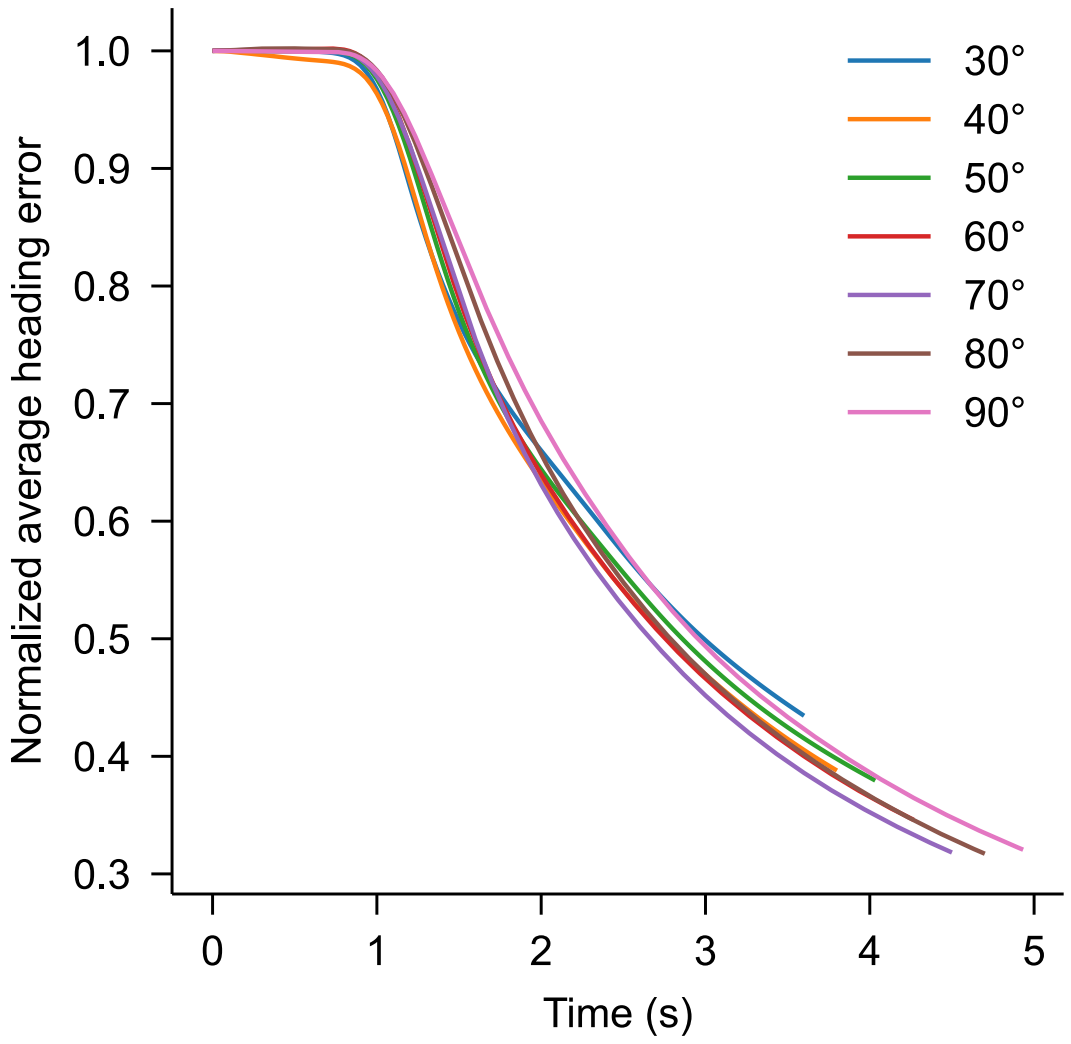

**Extended Data Fig. 5: Example path taken by a participant during personalization of audio feedback.** The participant performed a turning and walking task in an open field. They were guided with a series of random turns between 60 and 120 degrees to stay within a 10 meter radius area. Participant movement was estimated by using visual-inertial odometry. Arrows on the path indicate the direction that the participant was facing.

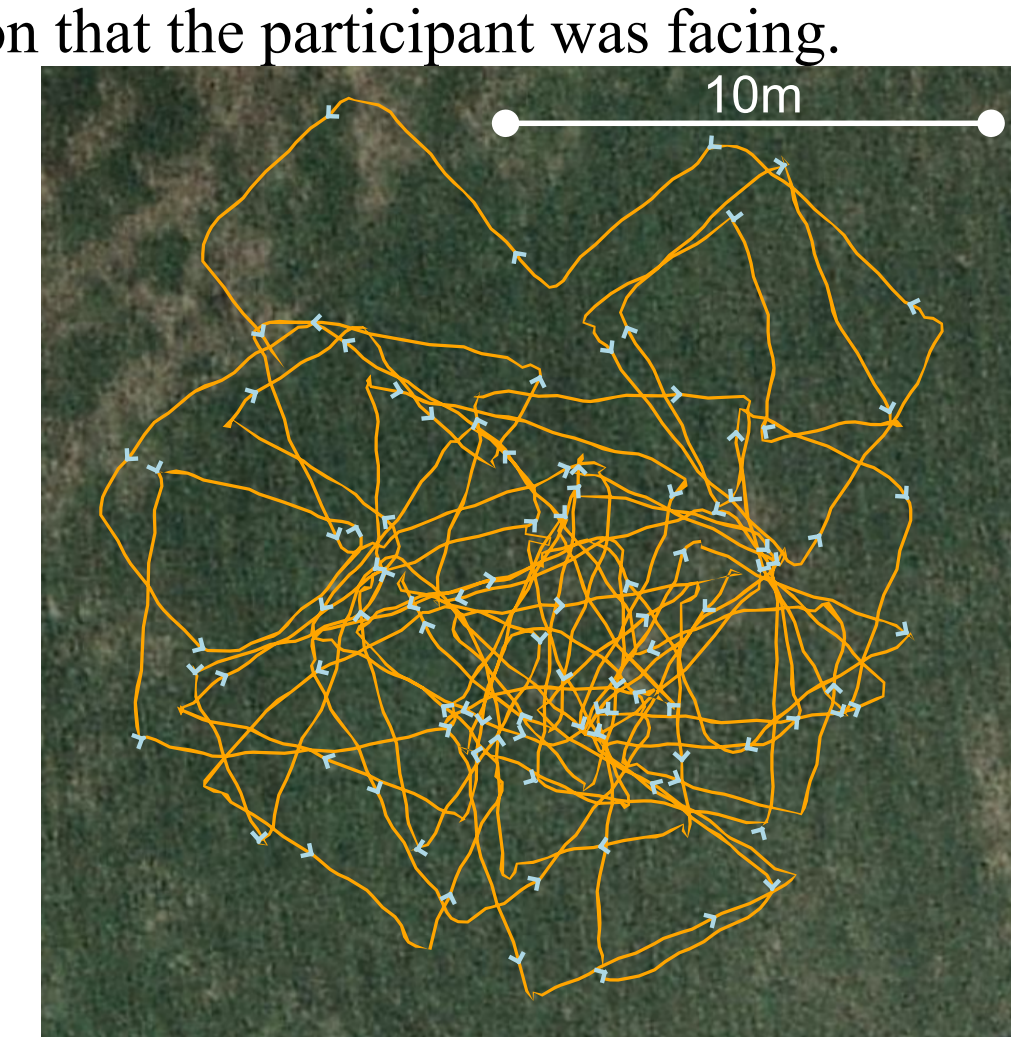

**Extended Data Fig. 6: Total number of verbal corrections across all participants.**

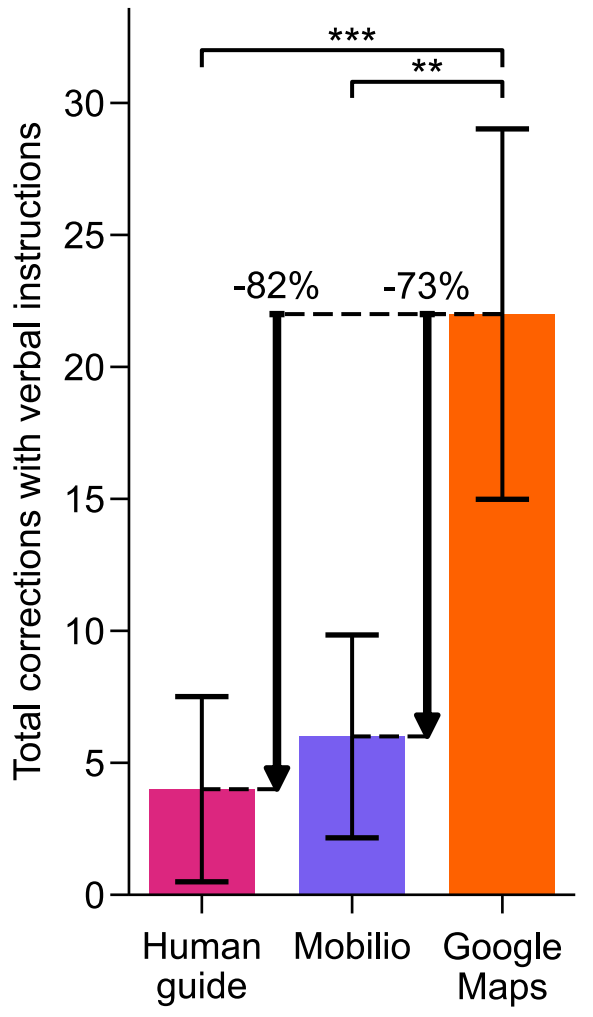

**Extended Data Fig. 7. Obstacle avoidance experiment. a,** Hallway obstacle course layout with movement paths for a representative participant using a white cane or Mobilio and a white cane. The two obstacles were pseudo-randomly placed for each course completion on the 0.6m placement intervals, at least 1.8m apart, with one on each side. **b,** Participants experienced fewer environmental contacts when using Mobilio with a white cane compared to using only a white cane (ANOVA, $n$ = 14, $P$ = 4.1e-6). **c,** Participants had a higher hallway completion time when using Mobilio (ANOVA, $n$ = 14, $P$ = 6.6e-3).

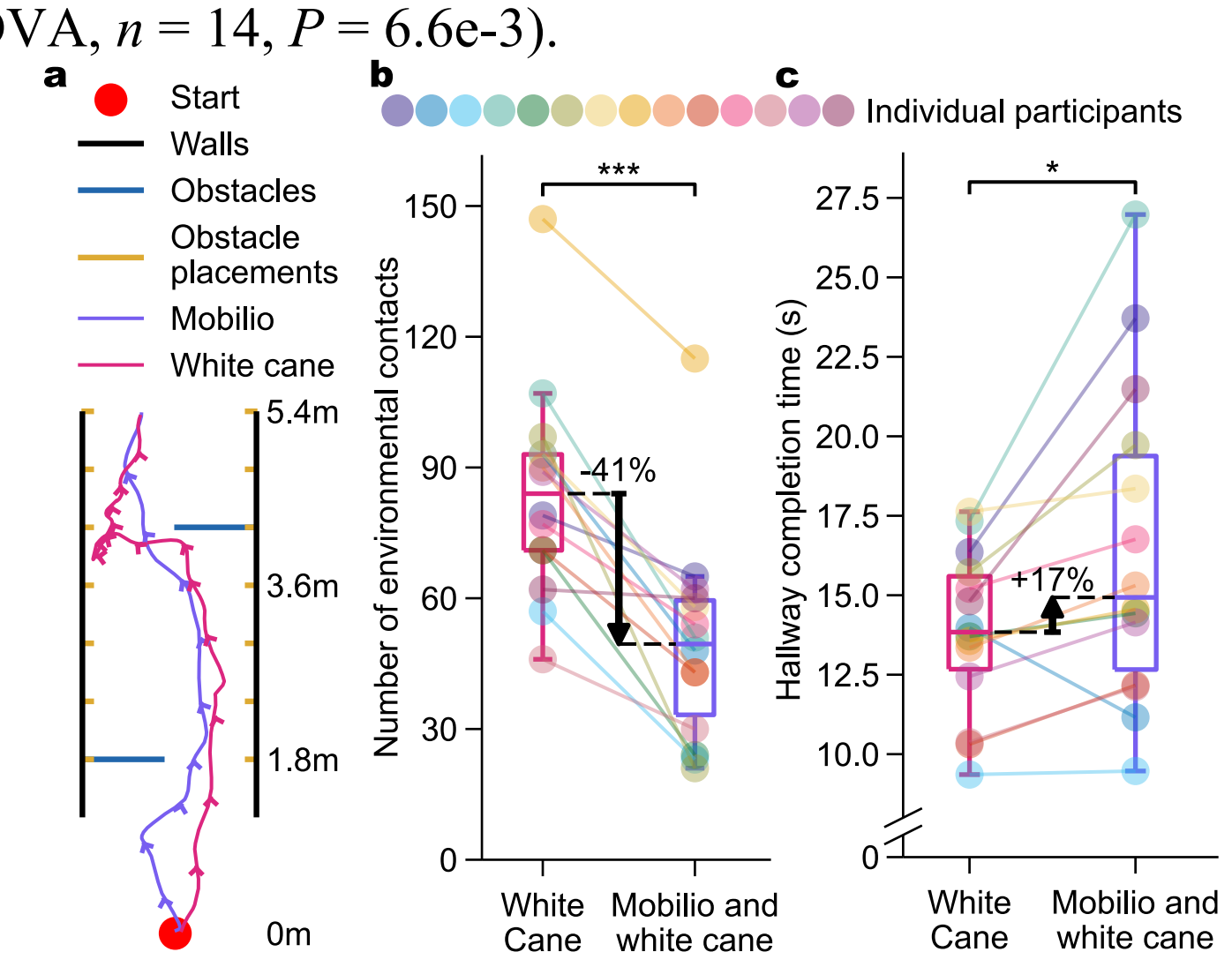

**Extended Data Fig. 8: LIDAR and monocular depth estimation evaluation during indoor walking.** A participant wearing the Mobilio smartphone walked towards five representative objects, with 5 repetitions per object, to obtain distance estimates from ground-truth motion capture, LIDAR, and monocular depth estimation. The participant started 5 meters away from the object and stopped a distance of 1 meter away from the object to approximate the distance they would contact the object if they were navigating with a white cane.

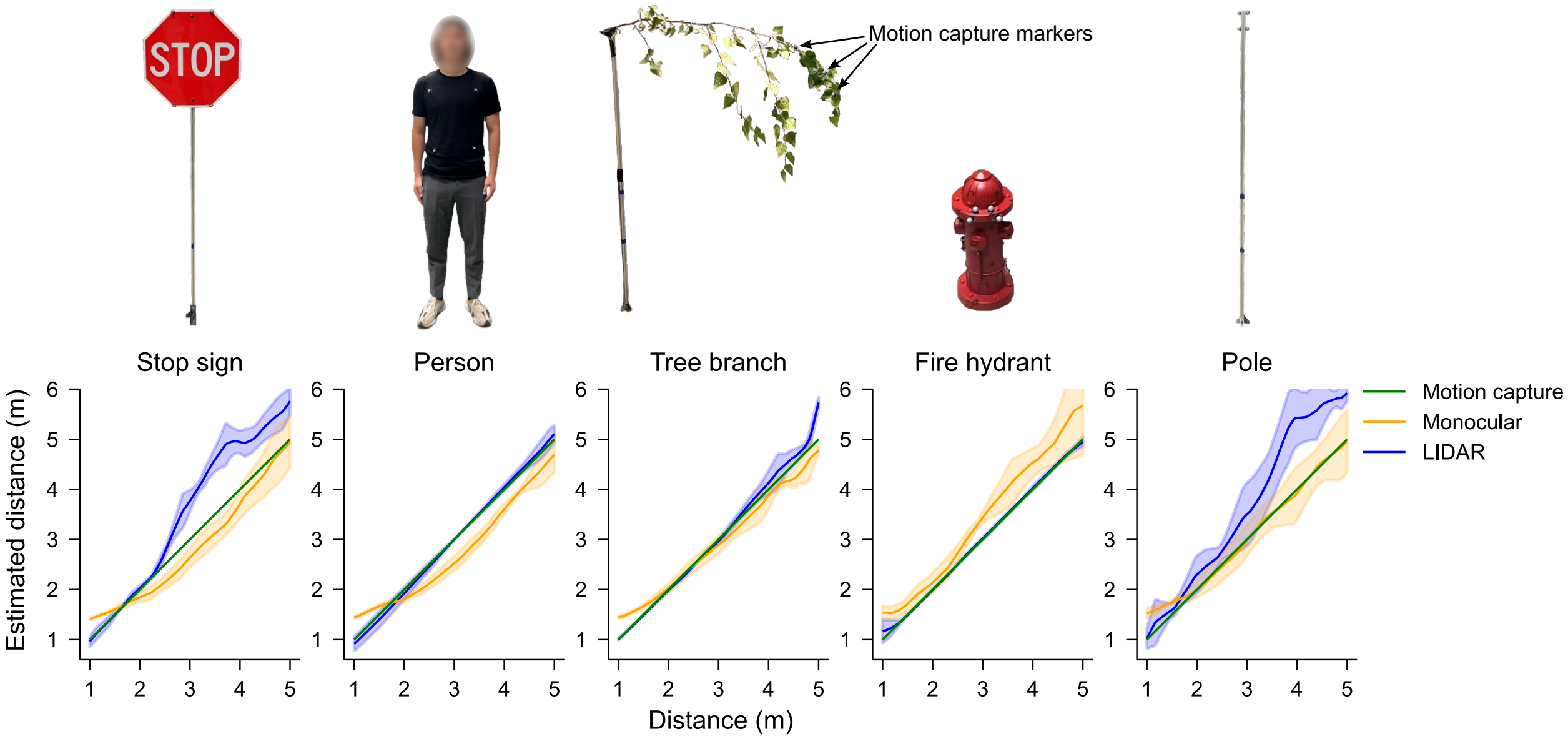

**Extended Data Fig. 9: Estimating scale and shift coefficients for monocular depth in an indoor lab space.** Each monocular depth value is multiplied by the scale coefficient and then has the shift coefficient added to estimate the true real-world distance to the environment. Darker colored points have a higher proportion of inliers from RANSAC, indicating higher confidence in the estimated scale and shift coefficients. Scale and shift coefficients were consistent for a representative trial where a person walks towards an obstacle. **a**, Scale decreased linearly as the distance to the obstacle decreased. **b**, Shift was approximately equal to 1 regardless of distance to the obstacle.

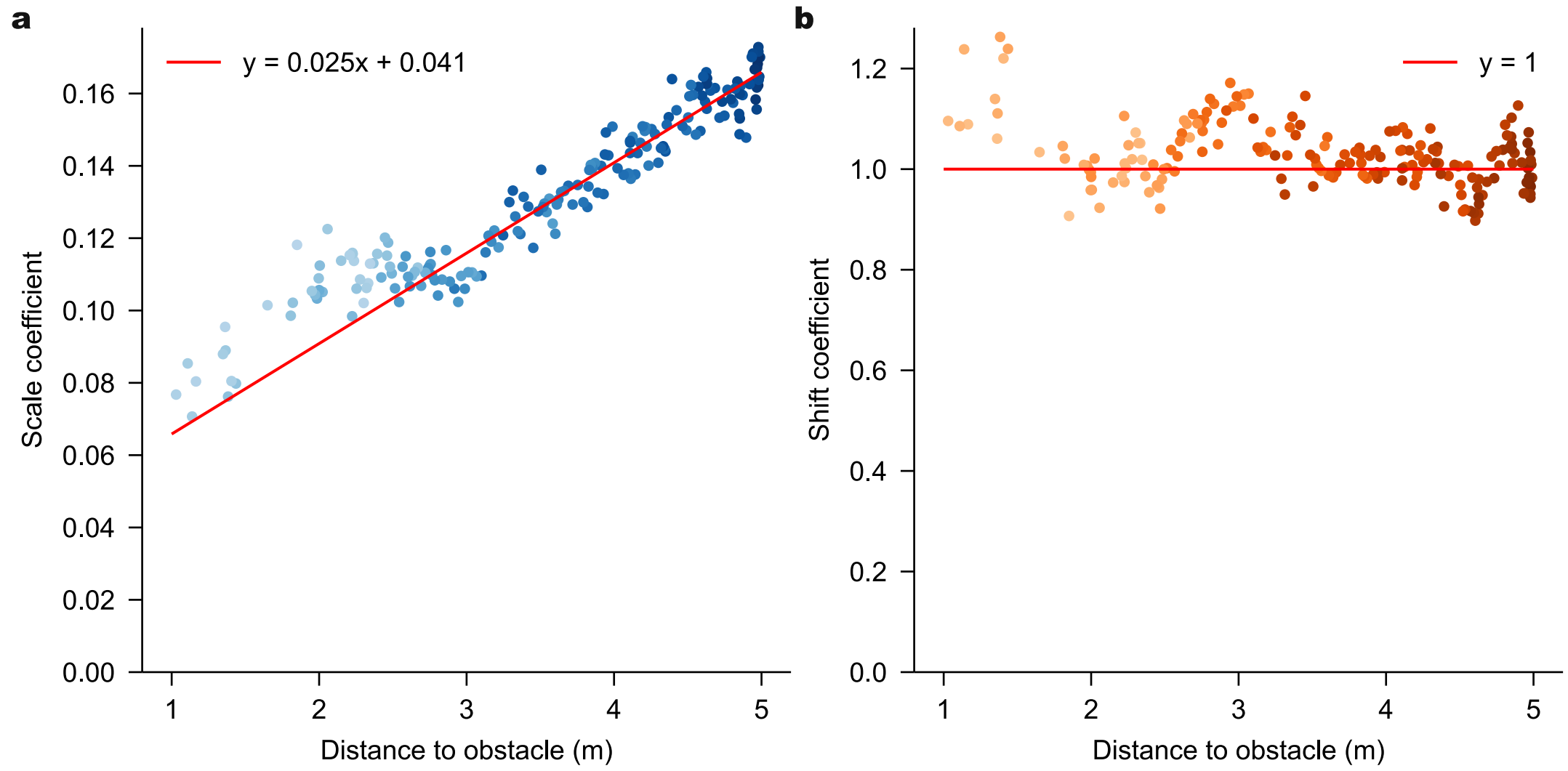

**Extended Data Fig. 10: Time-to-learn pilot experiment. a,** We evaluated the time it took participants to learn to use the audio feedback system of Mobilio in a turn-in-place experiment prior to the validation of Mobilio with participants with BVI. Ten participants without BVI (n = 10; 5 men and 5 women; age, 23.9 ± 1.51 yr; body mass, 69.2 ± 9.67 kg; height, 1.73 ± 0.06 m) were blindfolded and instructed to turn in place to face a series of randomized desired headings provided by audio feedback. The experiment consisted of repeating blocks of training trials and evaluation trials. **b,** Each trial tested three different target headings. During the training trials, audio feedback was personalized to each participant with CMA, using the same personalization approach described in this study for participants with BVI. During evaluation trials, participants received a pre-defined set of audio parameters to evaluate their performance at fixed intervals throughout the training process. **c,** This learning was modeled by fitting an exponential curve to the normalized heading errors during evaluation for all participants throughout the training process. The normalized average heading error for a trial is equal to the average heading error over the duration of the trial divided by the heading error at the start of the trial. Participants reached 86% of this steady-state exponential asymptote at 29 minutes, 90% at 34 minutes, and 95% at 44 minutes of training.

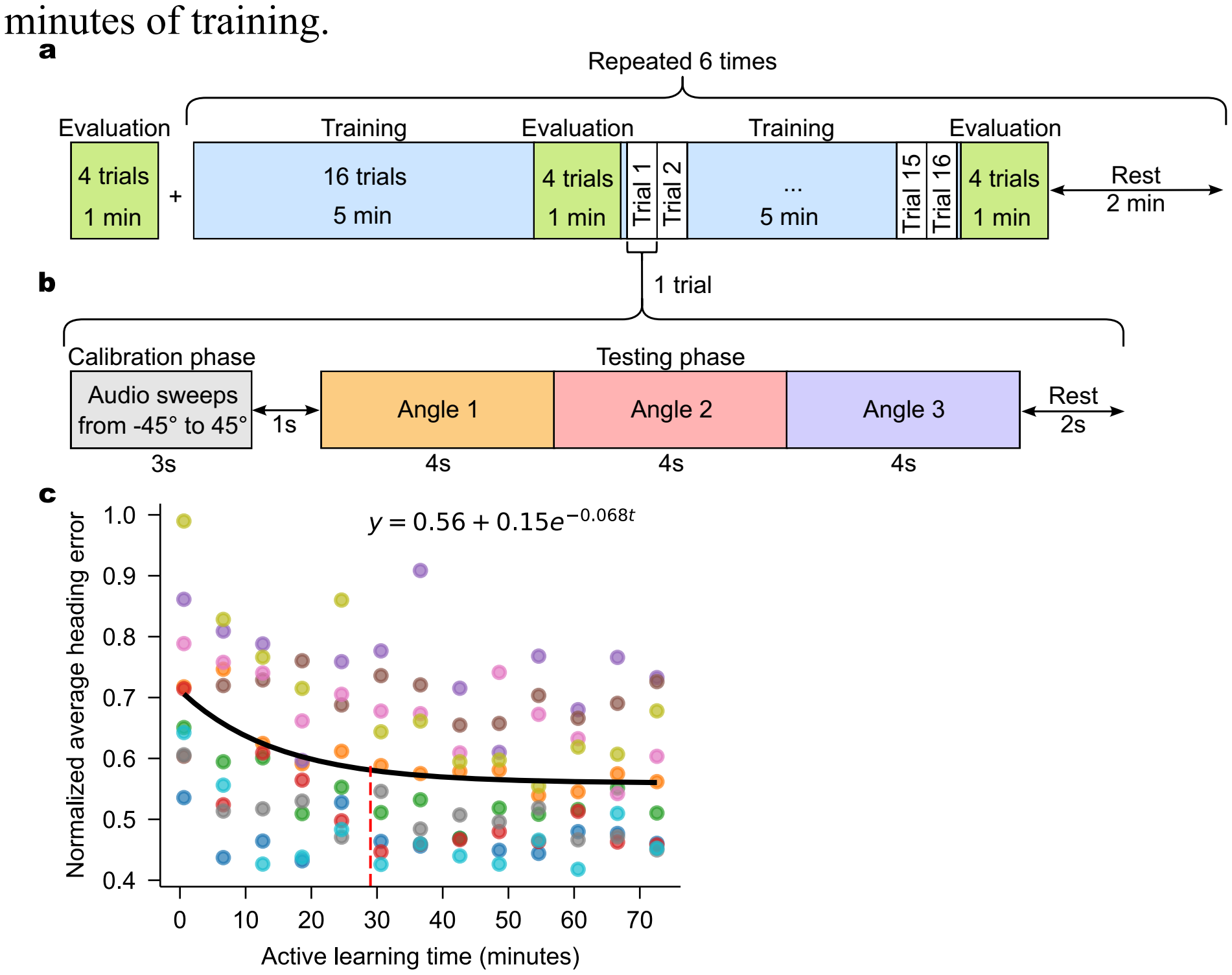

**Extended Data Fig. 11: Example participant routes requiring verbal corrections during community path navigation.** Arrows show the direction the participant is facing. Color indicates progress along the route snippet, with purple at the start and yellow at the end. **a,** A participant using Google Maps turns the wrong way on the first turn and walks past the second turn. Both times the research team provided them with verbal corrections after they walked in the incorrect direction for 10 seconds to allow them to continue making progress on the route. When using Mobilio, the same participant briefly walks past the second turn but corrects their trajectory with the help of continuous audio feedback without requiring verbal correction. **b,** A participant using Google Maps attempts to turn early due to receiving a turn instruction before they have reached the turn and becomes disoriented, requiring a verbal correction.

**Extended Data Table 1: Survey results on the necessary features for a navigation aid.** Survey respondents with BVI ($n$ = 112) completed this question without prior knowledge of Mobilio.

| **Which of these features are essential for an electronic travel aid to have in order to be useful to you? Check all that apply.** | **Count** | **Percentage** |
|---|---|---|
| Alerts me when I might run into an obstacle | 88 | 79% |
| Helps keep me on route to my destination | 83 | 74% |
| Can use GPS to reach my destination | 68 | 61% |
| Small enough to fit in a purse or backpack when I'm not using it | 68 | 61% |
| Long battery life | 62 | 55% |
| Lightweight | 60 | 54% |
| Does not require earphones to use | 46 | 41% |
| Something that I can wear | 40 | 36% |
| Something that can attach to my white cane | 29 | 26% |

**Extended Data Table 2: Survey results on why people with BVI do not use navigation aids.** Survey respondents with BVI ($n$ = 112) completed this question without prior knowledge of Mobilio.

| **If you have heard of electronic travel aids before, why don't you use one now? Check all that apply.** | **Count** | **Percentage** |
|---|---|---|
| They are too expensive | 39 | 35% |
| I do use one | 28 | 25% |
| I don't think they work well | 27 | 24% |
| I don't know where to buy them | 25 | 22% |
| I don't think they would help me | 16 | 14% |
| I prefer to use something else | 15 | 13% |
| I think they are too hard to use | 12 | 11% |
| I am not interested in buying a new device | 2 | 2% |

**Extended Data Table 3: Survey results on what people with BVI use to navigate outside.** Surveyed users respondents with BVI ($n$ = 112) completed this question without prior knowledge of Mobilio.

| **Do you use anything else to help you get from one place to another? List everything used.** | **Count** | **Percentage** |
|---|---|---|
| White cane | 92 | 82% |
| Smartphone GPS navigation apps (e.g. Google Maps) | 50 | 45% |
| Electronic travel aids (e.g. WeWalk) | 28 | 25% |
| Guide dog | 25 | 22% |
| Telescope, monoculars, or another device with a lens | 17 | 15% |
| Smartphone visual interpreter apps (e.g. Be My Eyes) | 16 | 14% |
| Smartphone apps - miscellaneous or unspecified | 16 | 14% |
| Rideshare or public transit apps (e.g. Uber) | 9 | 8% |

**Extended Data Table 4: Randomized condition ordering for the outdoor community path evaluations.** Conditions were randomized for each participant and presented in a double-reversal order. Participants used a white cane at all times.

| Subject | Condition 1 | Condition 2 | Condition 3 | Condition 4 | Condition 5 | Condition 6 |
|---|---|---|---|---|---|---|
| 1 | Google Maps | Human guide | Mobilio | Mobilio | Human guide | Google Maps |
| 2 | Mobilio | Google Maps | Human guide | Human guide | Google Maps | Mobilio |
| 3 | Human guide | Mobilio | Google Maps | Google Maps | Mobilio | Human guide |
| 4 | Google Maps | Human guide | Mobilio | Mobilio | Human guide | Google Maps |
| 5 | Human guide | Mobilio | Google Maps | Google Maps | Mobilio | Human guide |
| 6 | Google Maps | Mobilio | Human guide | Human guide | Mobilio | Google Maps |
| 7 | Human guide | Mobilio | Google Maps | Google Maps | Mobilio | Human guide |
| 8 | Human guide | Mobilio | Google Maps | Google Maps | Mobilio | Human guide |
| 9 | Human guide | Google Maps | Mobilio | Mobilio | Google Maps | Human guide |
| 10 | Google Maps | Mobilio | Human guide | Human guide | Mobilio | Google Maps |
| 11 | Mobilio | Google Maps | Human guide | Human guide | Google Maps | Mobilio |
| 12 | Human guide | Mobilio | Google Maps | Google Maps | Mobilio | Human guide |
| 13 | Mobilio | Google Maps | Human guide | Human guide | Google Maps | Mobilio |
| 14 | Google Maps | Mobilio | Human guide | Human guide | Mobilio | Google Maps |

**Extended Data Table 5: Randomized condition ordering for the hallway obstacle course evaluations.** Conditions were randomized for each participant and presented in a double-reversal order. Participants used a white cane at all times either with or without assistance from Mobilio.

| **Subject** | **Condition 1** | **Condition 2** | **Condition 3** | **Condition 4** |
|---|---|---|---|---|
| 1 | Mobilio | White cane | White cane | Mobilio |
| 2 | White cane | Mobilio | Mobilio | White cane |
| 3 | White cane | Mobilio | Mobilio | White cane |
| 4 | White cane | Mobilio | Mobilio | White cane |
| 5 | White cane | Mobilio | Mobilio | White cane |
| 6 | White cane | Mobilio | Mobilio | White cane |
| 7 | Mobilio | White cane | White cane | Mobilio |
| 8 | Mobilio | White cane | White cane | Mobilio |
| 9 | Mobilio | White cane | White cane | Mobilio |
| 10 | White cane | Mobilio | Mobilio | White cane |
| 11 | Mobilio | White cane | White cane | Mobilio |
| 12 | Mobilio | White cane | White cane | Mobilio |
| 13 | Mobilio | White cane | White cane | Mobilio |
| 14 | Mobilio | White cane | White cane | Mobilio |

**Extended Data Table 6: Participant survey results on the usability of Mobilio.**
The System Usability Scale was used to evaluate the usability of Mobilio. Participants ($n$ = 14) completed this survey verbally after completing all experiments. Mobilio was in the 84th percentile of a distribution of 166 previous usability studies using SUS questionnaires.

| **Question text (5 = Strongly Agree, 4 = Somewhat Agree, 3 = Neither Agree nor Disagree, 2 = Somewhat Disagree, 1 = Strongly Disagree)** | **Mobilio (Mean ± SD)** |
|---|---|
| I think that I would like to use this system frequently. | 4.0 ± 0.76 |
| I found the system unnecessarily complex. | 1.93 ± 1.1 |
| I thought the system was easy to use. | 4.64 ± 0.48 |
| I think that I would need the support of a technical person to be able to use this system. | 1.86 ± 1.25 |
| I found the various functions in this system were well integrated. | 4.0 ± 0.85 |
| I thought there was too much inconsistency in this system. | 2.5 ± 1.05 |
| I would imagine that most people would learn to use this system very quickly. | 4.07 ± 0.96 |
| I found the system very cumbersome to use. | 1.86 ± 0.83 |
| I felt very confident using the system. | 4.07 ± 0.8 |
| I needed to learn a lot of things before I could get going with this system. | 1.21 ± 0.41 |
| **Total usability score (out of 100)** | **78.57 ± 6.98** |

**Extended Data Table 7: Participant survey results on the perceived workload of Mobilio.** Questions from the NASA Task Load Index were used, with the exception of temporal demand because participants were directed to walk at a pace comfortable for them. Participants ($n$ = 14) completed this survey verbally after completing all experiments. Mobilio, white cane, and Google Maps were in the 10th, 12th, and 40th percentiles in a distribution of 1173 aggregated workload scores reported across 237 publications.

| **Question text (5 = Very High, 4 = Somewhat High, 3 = Neither High or Low, 2 = Somewhat Low, 1 = Very Low)** | **Mobilio (Mean ± SD)** | **White cane (Mean ± SD)** | **Google Maps (Mean ± SD)** |
|---|---|---|---|
| Mental demand: How mentally demanding was using the system? | 2.93 ± 1.16 | 2.71 ± 1.58 | 3.0 ± 1.41 |
| Physical demand: How physically demanding was the system? | 1.71 ± 1.16 | 2.29 ± 1.33 | 2.14 ± 1.25 |
| Effort: How hard did you have to work to accomplish your level of performance? | 2.21 ± 1.32 | 2.29 ± 1.39 | 3.21 ± 1.15 |
| Frustration: How insecure, discouraged, irritated, stressed, and annoyed were you? | 1.64 ± 0.97 | 1.71 ± 0.96 | 2.79 ± 1.47 |
| **Question text (5 = Excellent, 4 = Good, 3 = Average, 2 = Poor, 1 = Terrible)** | | | |
| Performance: How successful were you in accomplishing what you were asked to do? | 4.36 ± 0.48 | 4.57 ± 0.62 | 3.14 ± 0.99 |
| **Total score (out of 100)** | **25.71 ± 11.85** | **27.14 ± 13.68** | **45.0 ± 14.15** |

**Extended Data Table 8: Participant responses to open ended survey questions about Mobilio.** Participants (*n* = 14) completed this survey verbally after completing all experiments.

| **Question 1: What are some of the system's best attributes?** |
|---|
| I like that it alerts you to take your turns (right or left turns). It also beeps in the right ear so you know to go right, same for the left ear. If I turn too far to the right then it centers me to the middle and helps me straighten out. |
| It's easy to use. It slowed me down a little bit so I paid more attention to what I was doing, and then I had the confidence to know where I was going. One of the really nice things compared to Gmaps was the constant feedback so you know where you're going, while Gmaps leaves you a little stranded. You feel way more connected when you have the beeping through the app. Overall It made things a lot less stressful for navigating. |
| It gave me direct feedback when I was walking. It was easy for me to keep aligned in the direction I was going. So if I was using it in an area I knew was clear of obstacles it would be easy for me to get around. |
| Audio feedback, clarity in terms of communication. |
| I like the audio feedback. I like that when you're headed in the right direction, it plays in both ears as opposed to just one ear so you know you're going the right way. |
| The technology is lightweight and easy-to-use since it's just an app. It's simple to understand and follow the cues. |
| Knowing which direction to walk in. Also a little less hitting obstacles with the phone & cane compared to just a cane. |
| It's simple, easy to use, and consistent. |
| It lets you know about potential obstacles before encountering them with your cane, it could be useful for staying in the middle of a pathway in an open space without any physical markers or landmarks. |
| I'm a crooked walker and it helps me straighten out with subtle cues, as opposed to just saying "turn right" or "turn left". |
| When someone is first learning a new route, this would be useful. If I was using the beeps on the first run when I didn't know there was a gravel path, that would have been better, because when I was using Google Maps I didn't trust that the gravel path would be correct. |
| Having additional feedback about obstacles beyond just what's felt with the cane and the spatial-ness of the audio. |
| It kept me more centered on the path and gave me more feedback than Google Maps did. I enjoyed the more constant feedback. |

| I liked the spatial audio and the different tones. |
|---|
| **Question 2: What are some of the system's worst attributes?** |
| It's a bit confusing. If you're turning, do you keep turning, or do you turn a little bit and wait, since the beep stays in one ear consistently for several seconds. When it's in the center (both ears) it's confusing to know what to do, do I keep going straight? Then it would jump to the left ear and jump to the right ear. The jumping back and forth gets a little confusing. |
| It would be nice if it told you a pre-emptive prompt for what the next turn will be so you know which side of the path to follow. For example, "in 200 feet take a left turn". I don't know if it would work entirely without a white cane in an urban environment. However, I can see it really being beneficial at home or at school (a relatively normal environment without stuff like stairs, curbs, potholes) where I don't want or need to use a cane. |
| It got the turns wrong, either it told me to turn after I passed the turn or it told me to turn 3 to 6 feet before the turn. |
| Not clear intuitively how audio pitch and rate interplay. Stereo audio understanding can be uncertain when it's playing in both ears, ideally when it's in the middle things become quiet. If you're in a noisy area or in a city and using this outside, it'll be information overload with too much audio. It depends on your surroundings. |
| Geolocation accuracy, sometimes it tells you to turn too early. Also, sometimes the direction flutters back and forth and takes a bit to settle down, making it harder to line up correctly. |
| Especially for the outdoors, hearing the continuous sound kind of interferes with the pleasant parts of being outside, and it competes with other sounds I need to listen to. |
| In the hallway, it was hard to use in such a small area, the difference between right and left wasn't always distinct. |
| The constant beeping makes it difficult to hear the environment nearby, even with open-ear headphones. |
| I was a little confused because the beeps were sometimes bouncing side to side. |
| For people who are sporadic and panic when traveling it would be bad. I noticed when I turn it would beep all over the place. For people who are calmer and wait for the beep to catch up then it's fine. |
| The noise of the beeping bothered me. |
| Maybe it could do more verbal cues such as "turn left" in addition to the beeps. |
| I got a little confused with the obstacles, especially when there's a narrow space and it starts beeping on both sides. |

| Question 3: Beyond the study, would you be interested in using this system for navigation? |
| --- |
| Yes, I would. I noticed a big difference when I was using it versus just relying on my cane to pay attention not to run into the obstacles. |
| Yes. For someone like me who's visually impaired but not blind, I don't want to always carry a cane. But with this, I don't have to carry a cane and can just use this to help navigate. For navigating, following phone GPS sucks, and this would be really useful for navigating. Also useful for pointing out the big things like fire hydrants and poles. |
| I would be interested in playing with it, although I still think there's some bugs to work out. |
| I would definitely like to evaluate it further in real situations like in my walk to the train station, which I can do with a cane, but I feel this would work better. |
| Depends on how expensive it is, but yeah, I'd try. |
| Yes. |
| Yes, definitely outdoors for sure, especially in big open areas without path guidelines (edges) to follow. Indoors it would be nice if it could indicate directions to go towards landmarks such as elevators, doors, or open seats. Also it would be nice if it gave a warning of drop-offs or maybe vehicles. |
| Yes. |
| Potentially. It could be an interesting tool to add to the toolbox. |
| Yes. |
| Yes, especially for walking around places like a park. Using Google Maps in places that need very subtle movement is hard, because being 10 feet off in one way or the other could get you lost. |
| I don't think so. |
| Yes, as another tool in the toolbox. |
| Yes. |

**Extended Data Table 9: Segmentation model comparison on Mapillary Vistas.** Our model achieves a mean intersection over union (IoU) of 51.90 and a total pixel accuracy of 96.42; the DeepLabv3+ model achieves mean IoU of 52.13 and a pixel accuracy of 90.80. The intersection over union (IoU) for each class measures the overlap between predicted area and the ground truth area. Pixel accuracy measures the percent of pixels that were correctly classified. The precision for a class measures the accuracy of model predictions for that class; recall measures how much of the true area for the class was correctly identified.

| **Class** | **IoU (Mobilio)** | **IoU (DeepLabv3+)** | **Precision (Mobilio)** | **Precision (DeepLabv3+)** | **Recall (Mobilio)** |
|---|---|---|---|---|---|
| Road | 90.35 | 85.22 | 94.83 | 93.19 | 95.03 |
| Curb | 45.95 | 56.77 | 66.33 | 76.08 | 59.93 |
| Curb cut | 13.52 | 14.45 | 17.68 | 17.39 | 36.50 |
| Sidewalk | 68.51 | 66.61 | 82.25 | 77.34 | 80.40 |
| Plain crosswalk | 36.64 | 27.24 | 63.14 | 33.59 | 46.62 |
| Zebra crosswalk | 66.21 | 59.00 | 81.78 | 64.09 | 77.67 |
| Covering | 34.45 | 39.2 (catch basin), 46.87 (manhole) | 57.88 | 48.28 (catch basin), 58.8 (manhole) | 45.99 |
| Terrain | 59.57 | 64.74 | 74.18 | 77.76 | 75.15 |